\documentclass[usenatbib,a4paper,times,fleqn]{mnras}
\usepackage{graphicx,natbib,times,amssymb,amsmath,pstricks}
\usepackage{aas_macros}
\usepackage{bm,url}
\usepackage{textcomp}
\usepackage{breakurl}
\usepackage{subfig}
\usepackage{mathtools}

\title[Eccentricity evolution during planet-disc interaction]{Eccentricity evolution during planet-disc interaction}

\author[Ragusa et al.]{\parbox{0.7\textwidth}{Enrico Ragusa$^{1}$\thanks{enrico.ragusa@unimi.it}, 
Giovanni Rosotti$^{2}$\thanks{rosotti@ast.cam.ac.uk},
Jean Teyssandier$^{3,4}$,
Richard Booth$^{2}$,
Cathie J. Clarke$^{2}$ \&
Giuseppe Lodato$^{1}$ 
} \\
$^{1}$Dipartimento di Fisica, Universit\`a Degli Studi di Milano, Via Celoria, 16, Milano, I-20133, Italy \\
$^{2}$Institute of Astronomy, University of Cambridge, Madingley Road, Cambridge, CB3 0HA, UK \\
$^{3}$Department of Applied Mathematics and Theoretical Physics, University of Cambridge, Cambridge CB3 0WA, UK\\
$^{4}$Cornell Center for Astrophysics, Planetary Science, Department of Astronomy, Cornell University, Ithaca, NY 14853, USA
}
\pagerange{\pageref{firstpage}--\pageref{lastpage}} \pubyear{2017}

\date{2017}
\begin{document}
\label{firstpage}
\bibliographystyle{mnras}
\maketitle

\begin{abstract}

During the process of planet formation, the planet-discs interactions might excite (or damp) the orbital eccentricity of the planet. In this paper, we present two long ($t\sim 3\times 10^5$ orbits) numerical simulations: (a) one (with a relatively light disc, $M_{\rm d}/M_{\rm p}=0.2$) where the eccentricity initially stalls before growing at later times and (b) one (with a more massive disc, $M_{\rm d}/M_{\rm p}=0.65$) with fast growth and a late decrease of the eccentricity. We recover the well-known result that a more massive disc promotes a faster initial growth of the planet eccentricity. However, at late times the planet eccentricity decreases in the massive disc case, but increases in the light disc case. Both simulations show periodic eccentricity oscillations superimposed on a growing/decreasing trend and a rapid transition between fast and slow pericentre precession. The peculiar and contrasting evolution of the eccentricity of both planet and disc in the two simulations can be understood by invoking a simple toy model where the disc is treated as a second point-like gravitating body, subject to secular planet-planet interaction and eccentricity pumping/damping provided by the disc.
We show how the counterintuitive result that the more massive simulation produces a lower planet eccentricity at late times can be understood in terms of the different ratios of the disc-to-planet angular momentum in the two simulations. In our interpretation, at late times the planet eccentricity can increase more in low-mass discs rather than in high-mass discs, contrary to previous claims in the literature.

\end{abstract}

\begin{keywords}
 protoplanetary discs -- planets and satellites: formation -- planet-disc interaction %
\end{keywords}

\section{Introduction}
\label{intro}
The discovery of a large number of extrasolar planets has shown that the average orbital eccentricity of planets in the Galaxy is higher than that observed in our solar system \citep{butler2006}. Two possible scenarios have been proposed during the past three decades in order to understand the origin of the orbital eccentricities observed in exoplanets. The first involves the interaction with other massive bodies in the system after the disc dispersal, in fact in a gas poor environment: for example the action of the Kozai-Lidov mechanism in the presence of massive planetary companions or a binary star companion (\citealp{naoz2016} and references therein) or planet-planet scattering \citep{ford1996,papaloizouTerq2001,ford2008,juric2008,mustill2017}.
The second involves the interaction of the planet at resonant locations with the protoplanetary disc in which it has formed (see \citealp{kley2012Rev} for a review).

In this second scenario, Lindblad resonances pump the planet eccentricity, while corotation resonances damp it \citep{goldreich1980}. For a planet embedded in the disc the effectiveness of corotation resonances in damping the eccentricity exceeds the pumping action of Lindblad resonances, implying that the planet-disc interaction tends to circularize the planet orbits \citep{cresswell2007,bitsch2010}. Nevertheless, gas depletion in the corotation region produced by a sufficiently massive planet (typically $M_{\rm p}> M_{\rm J}$) might lead to the growth of the eccentricity. The works that found a growth in the planet eccentricity can be classified into two macro-categories. Firstly there are those that do not require an initial planet eccentricity, and those that instead do require a minimum amount of planet eccentricity to allow its further growth. 

In the first case (initial planetary eccentricity $e_{{\rm p},0}=0$), the planet needs to carve a gap which is deep enough to ensure that the disc torque on the planet is dominated by the contribution of the outer Lindblad resonance 1:3 \citep{papaloizou2001,dunhill2013,bitsch2013}. The planet eccentricity grows for $M_{\rm p}\gtrsim 5-10 M_J$, consistent with the predictions by \citet{lubow1991}.

In the second case (growth with initial planetary eccentricity $e_{p,0}\neq 0$), even planets with masses as small as $M_{\rm p}\gtrsim 1M_J$ have been found to be able to produce a saturation of co-orbital and corotation torque allowing the growth of the planet eccentricity provided the initial eccentricity $e_{p,0}>0.01$ (\citealp{dangelo2006}; \citealp{duffell2015}). These findings are in line with the theoretical production of \citet{ogilvie2003} and \citet{goldreich2003}. 

This excitation mechanisms apparently stop when the planet eccentricity reaches values comparable to the disc aspect-ratio ($e\sim H/R$) for two main reasons \citep{duffell2015}: first, because the epicyclic motion becomes increasingly supersonic for growing eccentricity, implying a weakening of Lindblad resonances responsible for the eccentricity pumping \citep{papaloizou2000}; second, because if the eccentricity is sufficiently high, the planet hits the cavity walls rapidly damping the eccentricity.
The mass of the disc and the density profile have also been shown to play a role in determining whether the planet eccentricity will grow. In particular, \citet{dunhill2013} found that, for sufficiently massive companions ($M_{\rm p}=25 \, M_{\rm J}$), the  eccentricity grows when the mass ratio between the companion and the disc is above a certain threshold and the density profile is such that the 1:3 Lindblad resonance dominates the overall torque exerted by the disc on the companion. 

The general conclusion of most of these works is that the planet-disc interaction is not able to provide planet eccentricity growth above the value $e_{\rm p}\gtrsim 0.15$ \citep{dangelo2006,muller2013,duffell2015,thun2017}. However, \citet{papaloizou2001} found for masses $M_{\rm p}\gtrsim 20 M_{\rm J}$ that the companion eccentricity might reach values of up to $e_{\rm p}\approx 0.25$. 

The exchange of angular momentum between the disc and the planet causes a growth also in the disc eccentricity \citep{goldreich1981}, even when the planet has a circular orbit (\citealp{papaloizou2001};\citealp{kley2006}; \citealp{teyssandier2016}; \citealp{teyssandier2017}). The disc reacts to the presence of a planet producing an eccentricity profile decreasing with radius. We report that the growth of the disc eccentricity has been proposed as a possible explanation of non-axisymmetric features \citep{ataiee2013,ragusa2017} observed in a large number of transition discs (\citealp{casassus2016}, for a review) as an alternative scenario to the widely invoked vortex hypothesis \citep{regaly2012,ataiee2013,lyra2013}. For completeness, numerical simulations in the context of binary black hole mergers has also revealed the formation of eccentric cavities with higher secondary-to-primary mass ratios \citep{armitage2005,shi2012,dorazio2013,farris2014,dorazio2016,ragusa2016}, with important consequences for the modulation of the accretion rate.

\citet{teyssandier2016} studied the normal modes solutions to the analytical equations ruling the eccentricity evolution in discs, thus making predictions about the disc eccentricity radial profiles in the presence of a planet. 

It is important to notice that, mostly on account of the high computational cost of these simulations, in the aforementioned works the evolution of the eccentricity has never been explored beyond $t\gtrsim 2\times 10^4 $ planet orbits \citep{thun2017}.

Motivated by the recent observation of CI Tau by \citet{krull2016}, \citet{rosotti2017} performed long timescales calculations ($\sim 10^5$ orbits) in order to study the role of planet-disc interaction in exciting hot-Jupiters' eccentricity. Their simulations showed very prominent secular oscillations of the eccentricity with periodicities $\gtrsim 10^4$ orbits, superimposed on a roughly linear growth starting at a time of $\sim 4 \times 10^4$ orbits (doubling the eccentricity from $\sim 0.04$ over $\sim 10^5$ orbits) after an apparent stalling of the eccentricity evolution. Similar oscillations have also been observed in other works \citep{duffell2015,muller2013,bitsch2013,dunhill2013,thun2017}.

Even though the results in \citet{rosotti2017} were not able to prove that planet-disc interaction might provide an effective mechanism to excite the eccentricities observed in hot-Jupiters, they showed clearly that the fate of the planetary eccentricity at late times cannot be determined a priori without performing simulations that cover significant fraction of the entire life time of the system.

In this paper we present two long term numerical simulation of the disc-planet evolution for two different disc masses.
The lower mass simulation is the same presented in \citet{rosotti2017} but integrated three times longer ($\sim 3\times 10^5$ orbits); the other has a disc mass that is a factor of three higher and is integrated for a similarly long time. We show that the initial behaviour of the planet eccentricity can be completely reversed at late times. Then we will give a physical interpretation of the peculiar evolution using a simplified toy model.

This paper is structured as follows. In section \ref{numsim} we present the numerical setup we used for our simulations. In section \ref{results} we present the results of our simulations. In section \ref{sec:interpretation} we discuss the results and introduce a simplified toy model to describe the evolution of the eccentricity. Section \ref{sec:toymodelinterp} provides an interpretation of the results in terms of the simplified toy model. In section \ref{sec:conclusions} we draw our conclusions.

\section{Numerical simulations}\label{numsim}

We perform two long timescale ($t\sim 3\times10^5$ orbits) 2D hydrodynamical simulations of a planet embedded in a gaseous disc orbiting a central star using \textsc{fargo3d} \citep{fargo2016} for two different disc masses. The simulations were run on GPUs (Nvidia Tesla K20), for a total wall clock time exceeding 6 months. We use open boundary conditions at the inner edge of the computational domain and closed at the outer one. We use a polar grid composed by $n_r=430$ radial cells between $R_{\rm in}=0.2$ and $R_{\rm out}=15$ with logarithmic spacing, and $n_\phi=580$ azimuthal cells. The outer radius $R_{\rm out}$ of the domain has been chosen to be sufficiently large to prevent the boundary conditions from affecting the dynamics; the propagation of eccentric perturbations do not reach radii $R\gtrsim 10$ (this can be noticed in Fig. \ref{fig:ecccol}, which will be discussed in the following sections).
 We use units in which $a_{\rm p,0}=1$, where $a_{\rm p,0}$ is the initial semi-major axis of the planet, and $GM_\ast=1$, where $M_\star$ is the mass of the star and $G$ the gravitational constant. The simulations cover a time of $t=3\times 10^5 t_{\rm orb}$, where $t_{\rm orb}=2\pi\Omega_{\rm p}^{-1}$ is the initial planet orbital period ($\Omega_{\rm p}$ Keplerian orbital frequency).

We vary the disc mass while keeping fixed the other simulation parameters. One simulation uses a disc-to-planet mass ratio $q=M_{\rm d}/M_{\rm p}=0.2$ while the other $q=0.65$, where $M_{\rm p}/M_\star=0.013$ is the planet mass and $M_{\rm d}$ is the total disc mass. We will refer henceforth to the case $q=0.2$ as the ``light'' case and the case $q=0.65$ as the ``massive'' case. To satisfy these conditions, the initial surface density distribution is a radial power-law of the type $\Sigma=\Sigma_0 R^{-p}$ with $p=0.3$ with the addition of an exponential taper at $R=5$; $\Sigma_{0,{\rm l}}=4.8\times 10^{-5}$ for the light case and $\Sigma_{0,{\rm m}}=15\times 10^{-5}$ for the massive case. 

The choice of the parameters for these simulations follows that used in \citet{rosotti2017} and is based on the best fit model of the disc surrounding the star CI Tau, where an eccentric $13\, M_{\rm J}$ hot-Jupiter has been found \citep{krull2016}.

We use a locally isothermal equation of state imposing a power-law radial temperature profile which provides a disc aspect-ratio of the type $h=H/R=0.036 R^{\ell}$ with $\ell=0.215$.

We use a \citet{shakura1973} viscosity prescription with $\alpha=10^{-3} R^{-0.63}$ in order to obtain a stationary accretion profile if the planet was not present. 
  
The planet is initially absent and its mass is progressively increased during the first $50$ orbits. During this period of time the planet is kept on a circular Keplerian orbit at $a_{\rm p}=1$, then its orbital parameters are left free to evolve under the action of the disc torque in order to allow the planet migration and eccentricity growth.

\section{Results}\label{results}

\begin{figure*}
\begin{center}
\includegraphics[width=0.49\textwidth]{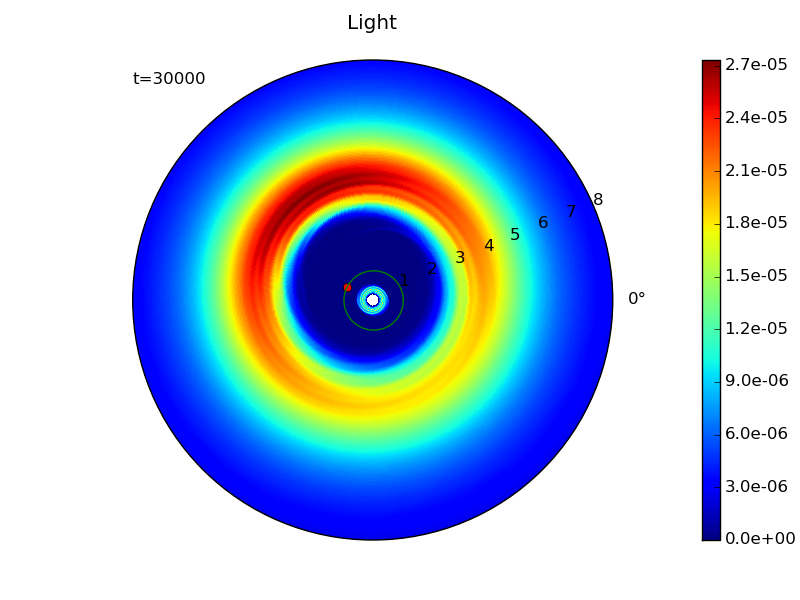}
\includegraphics[width=0.49\textwidth]{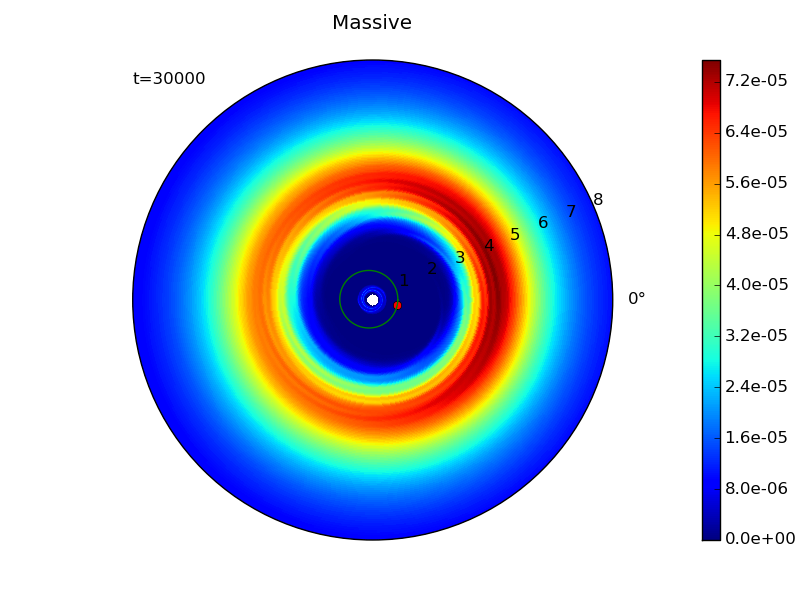}
\caption{Density colour-plot for light (left panel) and massive (right panel) case at $t=3\times 10^4 \,t_{\rm orb}$. The numbers indicate different radii. Note the formation of an eccentric cavity characterized by an horseshoe feature at its apocentre, consistent with the theoretical predictions about the density structure in eccentric discs.}
\label{fig:denscol}
\end{center}
\end{figure*}

\begin{figure*}
\begin{center}
\includegraphics[width=0.49\textwidth]{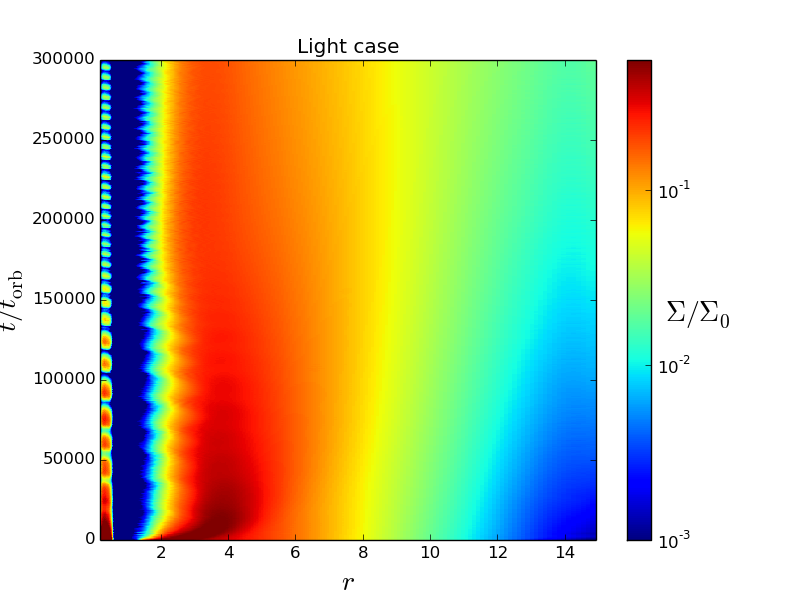}
\includegraphics[width=0.49\textwidth]{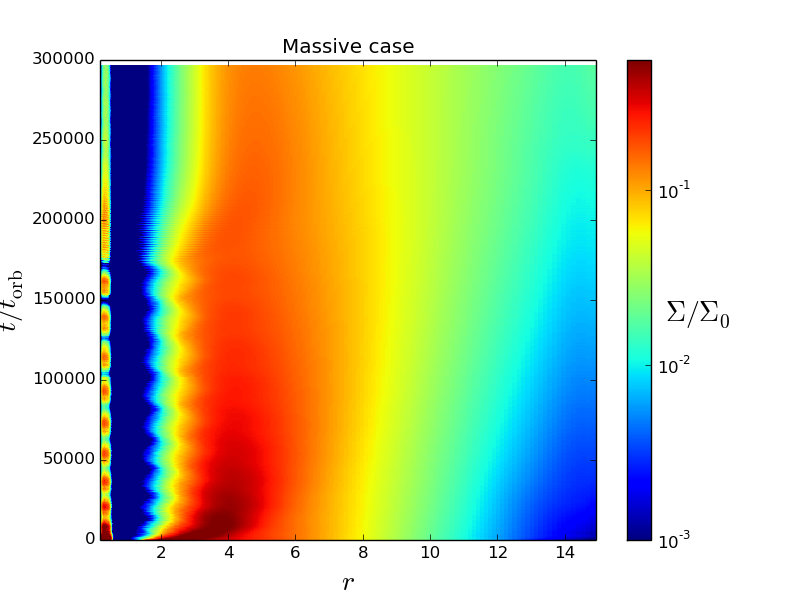}
\caption{Density radial profiles $\Sigma/\Sigma_0$, obtained through azimuthal average, for light (left panel) and massive (right panel) case as a function of radius (x-axis) and time (y-axis), different colours represent different values of density. }
\label{fig:denscolTIME}
\end{center}
\end{figure*}

\label{sec:results}

In Fig. \ref{fig:denscol} two colour maps of the disc surface density at $t=3\times10^4\,t_{\rm orb}$ are shown, both for the light and the massive disc case. The presence of an eccentric cavity and of a crescent shaped overdense feature at the apocentre of the cavity can be clearly noticed, consistent with the density perturbation expected for an eccentric disc (\citealp{teyssandier2016}, see their eq. A31). We also report that this type of features, induced by the presence of a planet or stellar companion, has been previously discussed in the literature \citep{ataiee2013,ragusa2017} to describe possible mechanisms producing the non-axisymmetric structures found in some transition discs in high resolution observations provided in the radio and NIR (\citealp{casassus2016}). In Fig. \ref{fig:denscolTIME} the time evolution of the disc density radial profile is shown. 

We used a Jacobi set of coordinates: thus the quantities related to the planet are computed in the reference frame of the star; while the quantities related to the disc are computed in the frame of the centre of mass (hereafter CM) of the system $M_\star+M_{\rm p}$. This peculiar set of coordinates is required since the disc orbits around the CM of the system. If computed in the star frame as the planet-related quantieties, the disc eccentricity would be non-vanishing at large radii.

As noted by \citet{ogilvie2001}, the eccentricity vector provides a useful tool to describe the values of eccentricity and pericentre phase of both the planet and the disc presented in this work. In our 2D case this reads
\begin{equation}
\bm e=-\frac{j}{GM} \hat{\bm u}_z\times\bm v- \hat{\bm u }_r,\label{eccvec}
\end{equation}
where $j$ is the modulus of the $z$-component of the the angular momentum vector per unit mass, $\bm v$ is the velocity vector of the planet (or of the disc fluid element considered when computing the disc eccentricity), $\hat{ \bm u}_z$ and $\hat{\bm u}_r$ are unit vectors pointing in the vertical and radial direction, respectively, and $M=M_\star+M_{\rm p}$. It can be shown that the modulus of $\bm e$ is the canonical expression for the orbital eccentricity $e$ 
\begin{equation}
e=\sqrt{1-\frac{j^2}{GMa}},
\end{equation}
where $a$ is the planet (or disc fluid element) semimajor-axis, and points in the pericentre direction.

The disc eccentricity is computed for each fluid element of the grid using Eq. \ref{eccvec}. The eccentricity radial profile is then obtained through an azimuthal average of grid cells at each radius. 

Beside the direct computation of the orbital eccentricity of the disc fluid elements, we can quantify the global amount of disc eccentricity using the angular momentum deficit (AMD), that is defined as follows
\begin{equation}
A_{\rm d}=\int \Sigma(R,\phi) \left[\sqrt{GMa(R,\phi)}- v_\phi(R,\phi) R\right]\, R dR \, d\phi,\label{AMDeq}
\end{equation}
where $\Sigma(R,\phi)$ is the disc surface density and $v_\phi(R,\phi)$ the azimuthal velocity and $a(R,\phi)$
\begin{equation}
a(R,\phi)=-\frac{1}{2}\frac{GM}{E(R,\phi)},
\end{equation}
where $E(R,\phi)$ is the mechanical energy per unit mass 
\begin{equation}
E(R,\phi)=-\frac{GM}{R}+\frac{1}{2}v^2(R,\phi),
\end{equation}
where $v(R,\phi)$ is the gas velocity map.

The AMD is the amount of angular momentum the disc is lacking in comparison with a situation where the gas orbits the CM of the system on circular orbits.
The same quantity is defined for the planet as follows
\begin{equation}
A_{\rm p}=J_{\rm circ,p}-J_{\rm p},\label{AMDeqP}
\end{equation}
where $J_{\rm circ,p}=M_{\rm p}\sqrt{GM_\star a_{\rm p}}$ is the angular momentum that the planet would have on a circular orbit with radius $a_{\rm p}$, $J_{\rm p}=M_{\rm p}v_{\phi,p}R_{\rm p}$ is the planet angular momentum where $v_{\phi,p}$ is the planet instantaneous azimuthal velocity and $R_{\rm p}$ is its separation from the central star. It can be shown that, for small eccentricities, Eq. \ref{AMDeqP} can be approximated by
\begin{equation}
A_{\rm p}\approx \frac{1}{2}e_{\rm p}^2J_{\rm circ,p}.\label{AMDappr}
\end{equation}

\subsection{Planet migration}

\begin{figure}
\begin{center}
\includegraphics[width=0.49\textwidth]{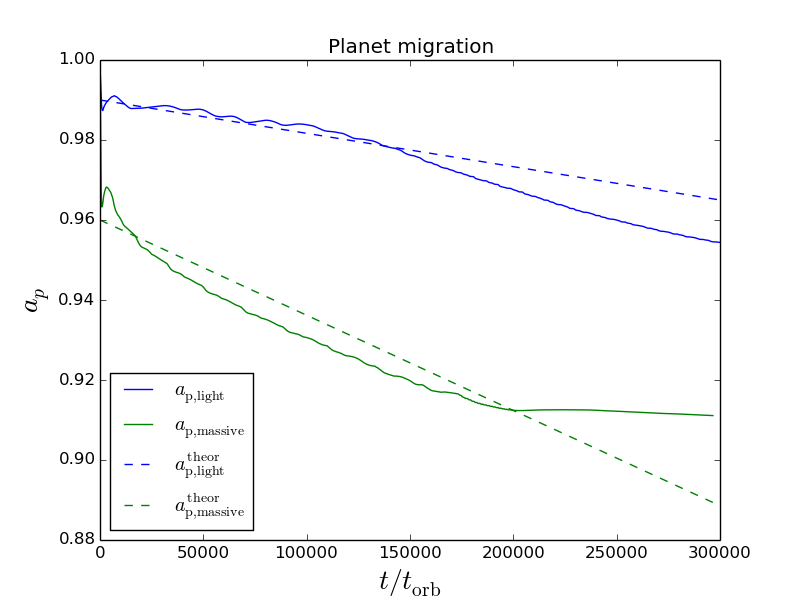}
\caption{Planet migration: $a_{\rm p}$ as a function of time for light (blue curve) and massive (green curve) case. The dashed lines are two lines with slope $\dot{a}_{\rm p}=( 100\, t_\nu)^{-1}$ for the light case and $\dot{a}_{\rm p}= (35 \,t_\nu)^{-1}$ for the massive one, which represents the theoretical migration rate predicted by Eq. (\ref{ttypeII}) for the two cases.}
\label{fig:migration}
\end{center}
\end{figure}

The planet-disc interaction also drives the migration of the planet, see Fig. \ref{fig:migration}.
This is actually a consequence of the conservation of the total angular momentum.
Indeed the total AMD $A_{\rm tot}=A_{\rm p}+A_{\rm d}$ can be written as
\begin{equation}
A_{\rm tot}=J_{\rm circ,p}+J_{\rm circ,d}-J_{\rm tot}\label{Atotmigr}
\end{equation}
where $J_{circ,\rm d}$ is the angular momentum of the disc if it was circular and $J_{\rm tot}$ is the total angular momentum of the system.
It follows straightforwardly that, in order to conserve the total angular momentum, any change in $A_{\rm tot}$ (which depends on the eccentricity of both the planet and the disc) in the simulations has to be accompanied by corresponding changes in $J_{\rm circ,d}+J_{\rm circ,p}$, i.e. varying the semi-major axis of the orbits both in the planet and in the gas.

We can compare the migration timescale $t_{\rm mig}=a_{\rm p}/\dot a_{\rm p}$ we observe in our simulation with the standard type II migration rate\footnote{It should be noticed that when the planet is left free to evolve it is completely embedded in the disc, and it spends the first $\approx 10^3$ orbits undergoing type I migration, migrating at a much faster rate. This produces the impression in Fig. \ref{fig:migration} that the initial $a_{\rm p,0}<1$. As soon as the cavity is cleared, it starts migrating at the slower type II rate.} \citep{syer1995,ivanov1999}
\begin{equation}
t_{\rm type II}=\frac{M_{\rm d}^{\rm local}+M_{\rm p}}{M_{\rm d}^{\rm local}}t_\nu\label{ttypeII}
\end{equation}
where $M_{\rm d}^{\rm local}=4\pi\Sigma(a_{\rm p})a_{\rm p}^2$ is approximately the unperturbed amount of disc material contained inside the orbit of the planet; $t_\nu$ is the viscous time-scale
\begin{equation}
t_\nu=(\alpha h^2\Omega_{\rm p})^{-1},
\end{equation}
where $h=H/R$ is the disc aspect ratio, $\alpha$ is the \citet{shakura1973} viscous parameter. Substituting the values from our simulations one gets $t_\nu\sim 1.2\times 10^5 t_{\rm orb}$ for both our setups, which is perfectly consistent with the damping timescale for the semi-major axis.
The timescales we obtain for type II migration from our simulations are $t_{\rm type II}^{\rm light}\sim 100\, t_\nu$ and $t_{\rm type II}^{\rm massive}\sim 35 \,t_\nu$, plotted as dashed lines in Fig. \ref{fig:migration}. We find that the ratio $t_{\rm type II}^{\rm light}/t_{\rm type II}^{\rm massive}\sim 2.85$ is perfectly consistent with the ratio one would expect from Eq. \ref{ttypeII} when comparing the migration rate of a planet embedded in two discs differing by factor 3 in the disc mass. The result is thus consistent with the classical Type II migration rate predicted by \citet{syer1995} and \citet{ivanov1999}, contrary to what have been found in the studies of \citet{durmann2015} and \citet{duffell2014}.

It should be noticed that the migration of the planet apparently stops at $t\sim 2\times 10^5$ orbits in the massive case. As we will see, this can be reasonably attributed to the rapid broadening of the cavity that can be noticed in Fig. \ref{fig:denscolTIME}. A larger cavity implies a clearing of material from the region where resonances mediate energy exchange between disc and planet. These structural changes also correspond to changes in the eccentricity evolution.
Interestingly, a similar behaviour has been previously observed in \citet{papaloizou2001} in which they observed a change of migration rate (and even direction) as a consequence of the disc structure evolution. In contrast, in the light case the migration accelerates at late times.
In the latter case the migration rate indeed appears to increase starting from $1.5 \times 10^5$ orbits , which (from Fig. \ref{fig:denscolTIME}) can be seen to coincide with the disc's inner edge moving slightly inwards.

\subsection{Eccentricity and pericentre phase evolution}\label{sec:eccevo}

\begin{figure*}
\begin{center}
\includegraphics[width=0.49\textwidth]{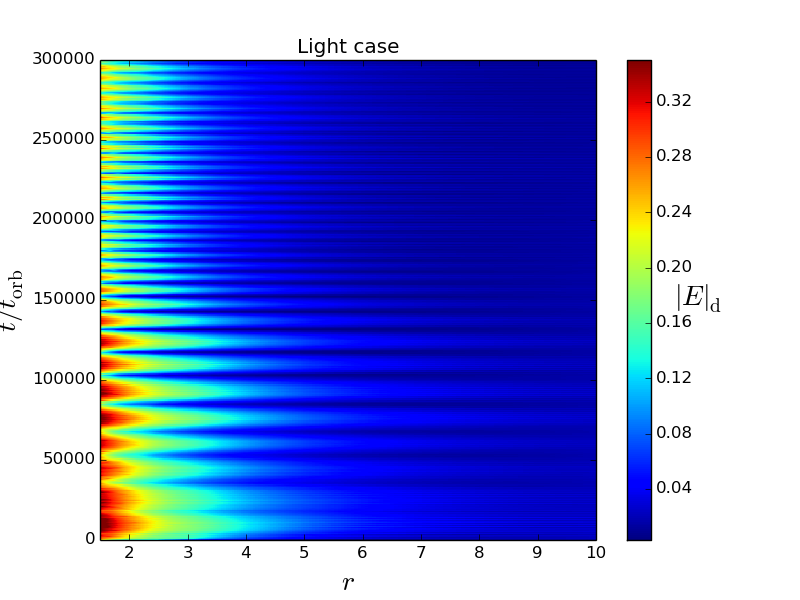}
\includegraphics[width=0.49\textwidth]{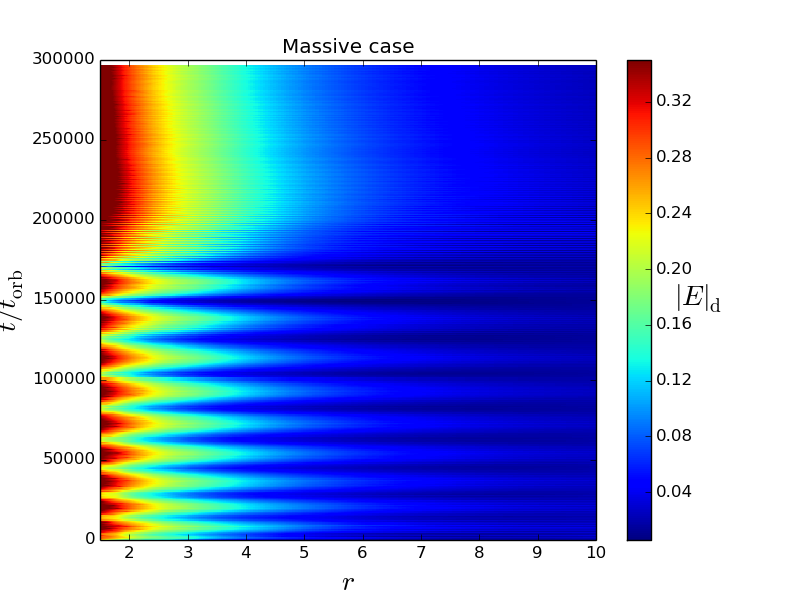}
\caption{Colour plots of the eccentricity (azimuthal average) as a function of time (y-axis) and radius (x-axis) for light (left panel) and massive (right panel) case.}
\label{fig:ecccol}
\end{center}
\end{figure*}

\begin{figure*}
\begin{center}
\includegraphics[width=0.49\textwidth]{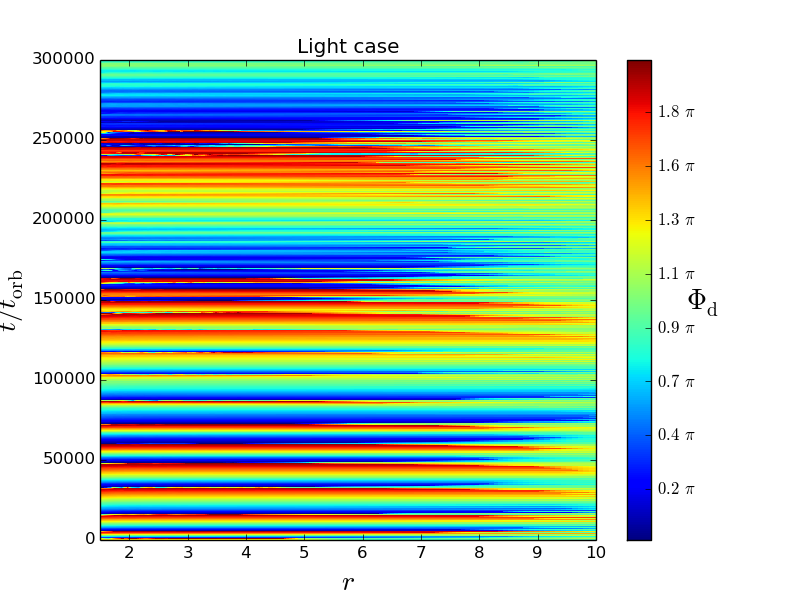}
\includegraphics[width=0.49\textwidth]{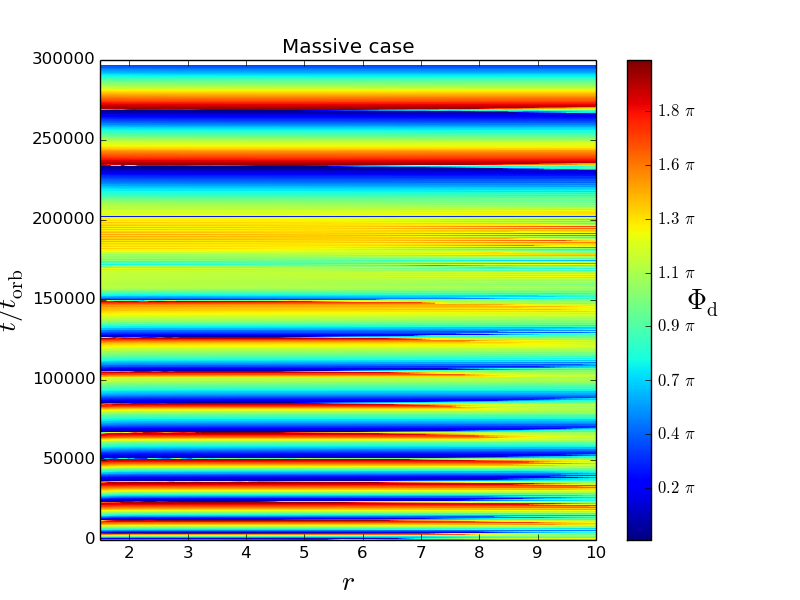}
\caption{Pericentre phase colour-plot as a function of time (y-axis) and radius (x-axis) for light (left panel) and massive (right panel. ) case. }
\label{fig:phasecol}
\end{center}
\end{figure*}

\begin{figure*}
\begin{center}
\includegraphics[width=0.49\textwidth]{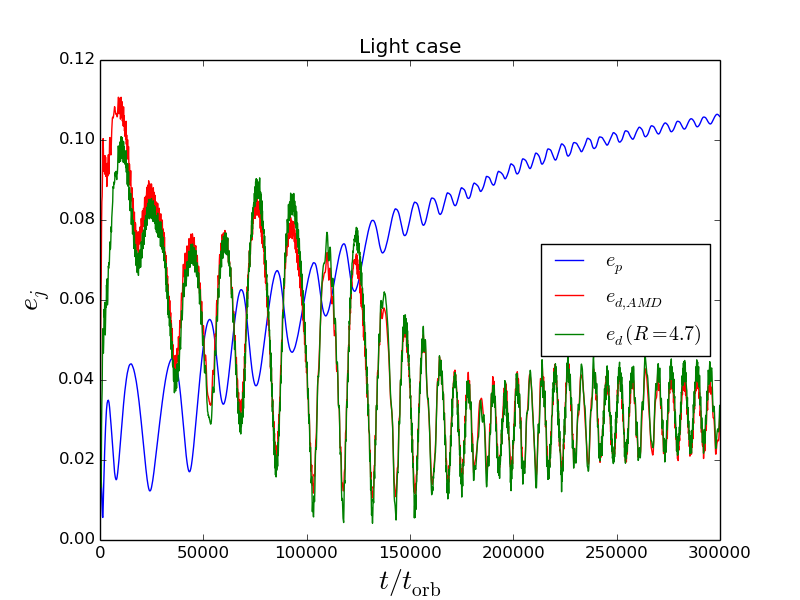}
\includegraphics[width=0.49\textwidth]{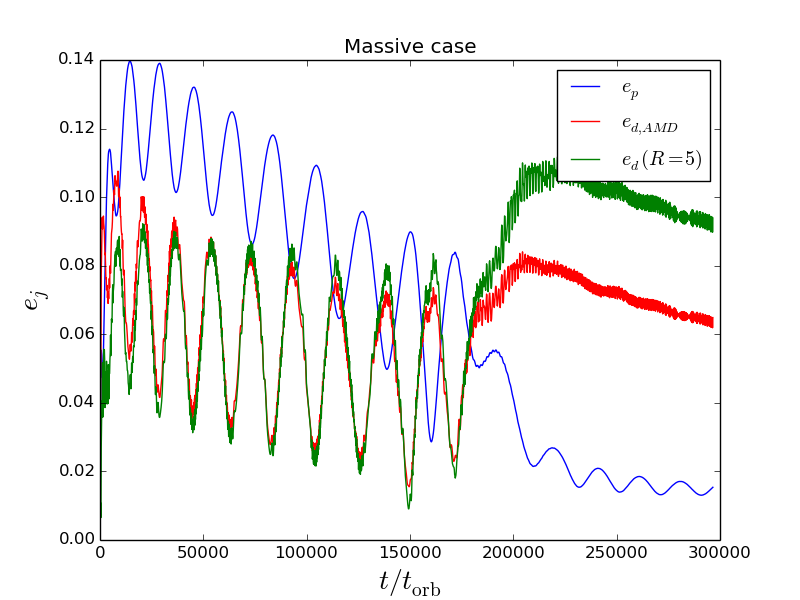}
\caption{Eccentricity $e$ as a function of time for light (left panel) and massive (right panel) case. The blue curve shows the planet eccentricity, the green curve the disc eccentricity at $R=4.7$ in the light case and at $R=5$ (azimuthal averages) in the massive one, while the red curve is a global measurement of the disc eccentricity starting from the AMD (see Sec. \ref{sec:eccevo}). The choice to use two different reference radii for the disc eccentricity is due to the slightly different size of the cavity in the two cases.}
\label{fig:ecc}
\end{center}
\end{figure*}

\begin{figure*}
\begin{center}
\includegraphics[width=0.49\textwidth]{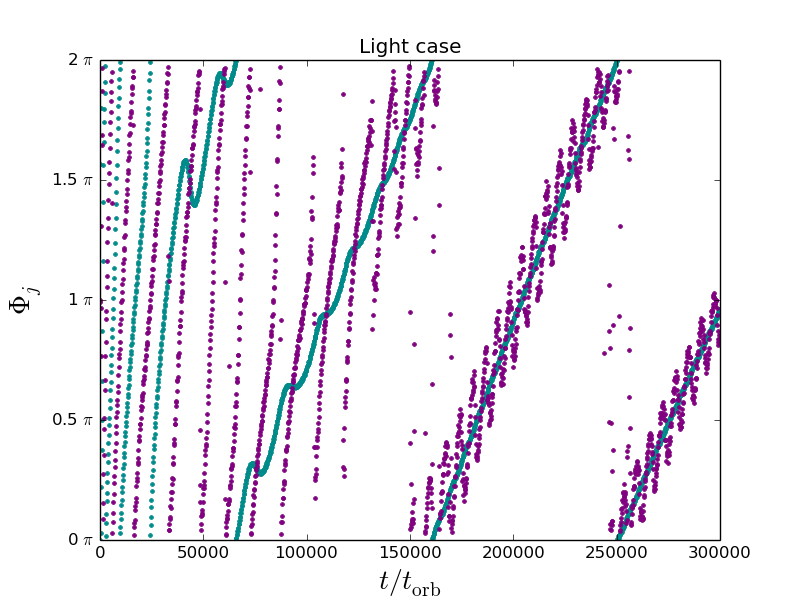}
\includegraphics[width=0.49\textwidth]{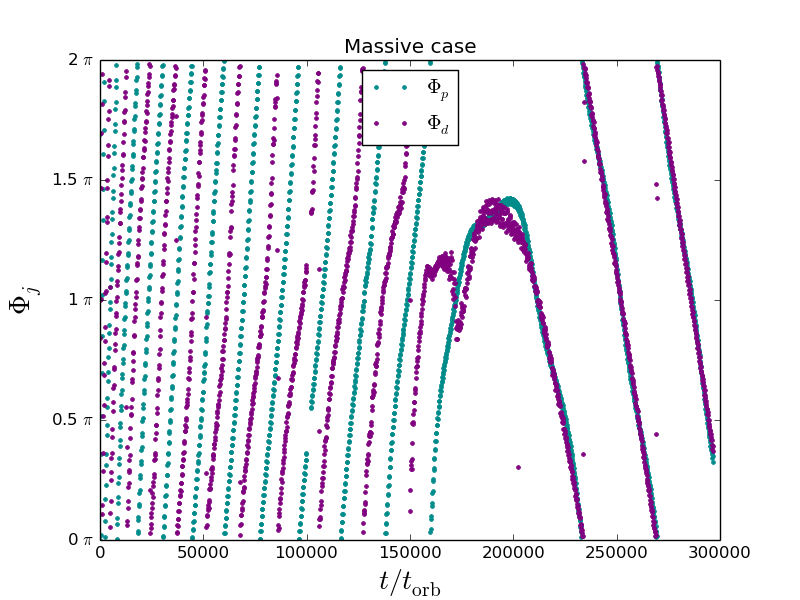}
\caption{Pericentre phase as a function of time for light (left panel) and massive (right panel) case (we remind that the value of the pericentre phase is constant throughout the entire disc domain). The cyan and violet curve represent planet and disc pericentre phase. During the first $\approx 4\times 10^4$ orbits both simulations show an anti-aligned precession ($|\Phi_{\rm p}-\Phi_{\rm d}|\approx 180^\circ$). After $\approx 4\times 10^4$ orbits in the light case the planet precession decouples from that of the disc, becoming much slower than the disc one. At very late times ($t\gtrsim 2\times 10^5$ orbits) also the disc precession rate slows down, and precesses along the planet in a pericentre aligned configuration. The massive case remain in the anti-aligned configuration for much longer, even though also in this case a transition toward the slowly precessing aligned configuration takes place after $t\approx 2\times 10^5$ orbits. In the massive case, the transition appears to be accompanied by a reversal of the precession rate, which becomes retrograde} 
\label{fig:phase}
\end{center}
\end{figure*}

In this section we will limit our discussion to the qualitative behaviour of the eccentricity evolution in the simulations; We postpone a possible modelling and interpretation of the results to the following sections.

The colour plots in Fig. \ref{fig:ecccol} and \ref{fig:phasecol} show the eccentricity (azimuthal average) and pericentre phase, respectively at different times ($y$-axis) and radii ($x$-axis) both for the light and massive disc case. 

It is interesting to note from these plots that the disc eccentricity evolution can be considered ``rigid'': in Fig. \ref{fig:ecccol}, for any fixed time, an increase in the eccentricity at small radii is reflected in an increase also at larger radii. The disc ``rigid'' behaviour is even more evident looking at Fig. \ref{fig:phasecol}, at fixed time, the pericentre phase is the same at all radii throughout the entire disc\footnote{For $R>8$ the eccentricity is almost negligible. The algorithm we used to compute the pericentre phase tends to attribute $\Phi_{\rm d}=180^\circ$ when $e\approx 0$ since it is not possible to attribute a pericentre in a circular orbit.}. Furthermore the radial profiles of the eccentricity and pericentre phase in Fig. \ref{fig:ecccol} and \ref{fig:phasecol} imply that the gas orbits are a set of nested, pericentre aligned eccentric orbits with an eccentricity profile decreasing with radius.

Fig. \ref{fig:ecc} and \ref{fig:phase} show the evolution of planet and disc eccentricity and their pericentre phases, respectively. The red and green curves in Fig. \ref{fig:ecc} represent two different ways to estimate the disc eccentricity: the red curve ($e_{\rm d,AMD}$) is computed inverting the approximate relationship between eccentricity and
AMD given in the case of the planet in Eq. (\ref{AMDappr}) and which yields  
\begin{equation}
e_{\rm d,AMD}=\sqrt{\frac{2A_{\rm d}}{J_{\rm d,circ}}},\label{redcurve}
\end{equation}
where $J_{\rm d,circ}$ is given by
\begin{equation}
J_{\rm d,circ}=\int \Sigma(R,\phi) \sqrt{GMa(R,\phi)}\, R dR\,d\phi;
\end{equation}
this approach allows us to give an estimate of the disc eccentricity relying on global disc quantities, in fact treating it as if it was a second planet.
We then notice that the density peak in the two simulations is located at $R\approx 4.7$ in the light case and at $R\approx 5$ in the massive one; the green curve represents the values of the eccentricity evolution at these radii.
The good agreement of the two curves tells us that the global behaviour of the disc is dominated by the values the eccentricity has at these radii, the discrepancy in the massive case between the red and green curve after $\approx 1.7\times 10^5$ orbits suggest that the reference radius for the eccentricity has migrated outward (consistently with the broadening of the cavity, Fig. \ref{fig:denscolTIME}).

Both simulations show that the planet and the disc exchange eccentricity through slow periodic (period $\Delta t\gtrsim 10^4$ orbits) anti-phased oscillations (a maximum in the planet curve correspond to a minimum in the disc one) superimposed on a series of roughly linearly growing and decreasing trends. It should be noted that the frequency of the oscillation is not constant throughout the entire length of the simulation, we will discuss more in detail this feature in Sec. \ref{sec:effectsdiscevo}.

Both the light and the massive case show a rapid exponential growth of the disc eccentricity (Fig. \ref{fig:ecc}) during the first stages of evolution ($t\lesssim 1.5\times10^4$ orbits) up to values $e_{\rm d}\sim 0.11$, then a slower decrease at later times. Interestingly, the maximum level of disc eccentricity achieved is the same for both simulations. This might suggest that some non-linear effects prevent the disc eccentricity to grow further. The planet eccentricity in the massive case has a similar behaviour: it grows fast in the beginning, attains a value $e_{\rm p}=0.14$ and starts decreasing at the same time as the disc eccentricity. The planet eccentricity in the light disc case in contrast has a completely different behaviour: its growth oscillates around $e_{\rm p}=0.025$ for $t\lesssim 4\times10^4$ orbits, but then at later times starts growing again at constant rate; at very late times ($t\gtrsim 2\times 10^5$ orbits) the planet eccentricity growth rate appears to slow down. 

The precession of the pericentre phase (Fig. \ref{fig:phase}) presents some very interesting features as well. Both the massive and the light disc cases show in the initial stages ($t\lesssim 4\times 10^4$ orbits) of the simulation an anti-phased precession of the planet-disc pericentre: the planet and disc pericentre precess at the same rate maintaining a phase difference $|\Phi_{\rm p}-\Phi_{\rm d}|\approx 180^\circ$. 

In the light case, after $\approx 4\times 10^4$ orbits, in correspondence with the beginning of the growing trend of the planet eccentricity, the planet pericentre phase starts to precess much more slowly than the disc one. After $t\gtrsim 1.5\times 10^5$ orbits also the disc transitions to a slower precession rate. When this condition is reached, the planet and the disc pericentre phases are aligned, precessing at the same slow rate. Some oscillations at a faster frequency in the disc pericentre phase can be noticed in this slow configuration.

In the massive case, the anti-aligned configuration (present in the light case just for $t\lesssim 4\times 10^4$ orbits) lasts for much longer: only at $t\gtrsim 2\times 10^5$ the transition toward the slower aligned precession rate appears to take place. However, in this case the transition appears to be accompanied by a reversal of the precession rate (which becomes retrograde), a significant slow down of the oscillation frequency of the eccentricities and also by a variation in the disc eccentricity radial profile (see the right panel of Fig. \ref{fig:ecccol}). 

\begin{figure*}
\begin{center}
\includegraphics[width=0.49\textwidth]{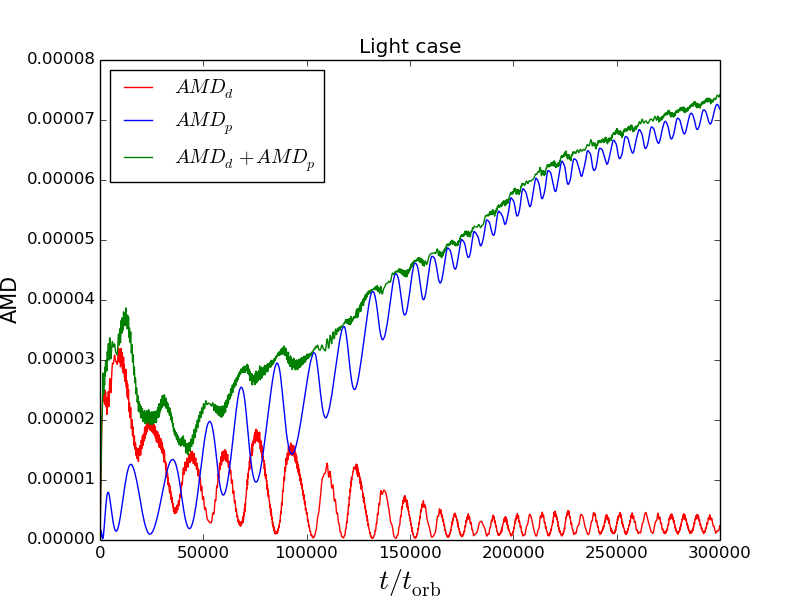}
\includegraphics[width=0.49\textwidth]{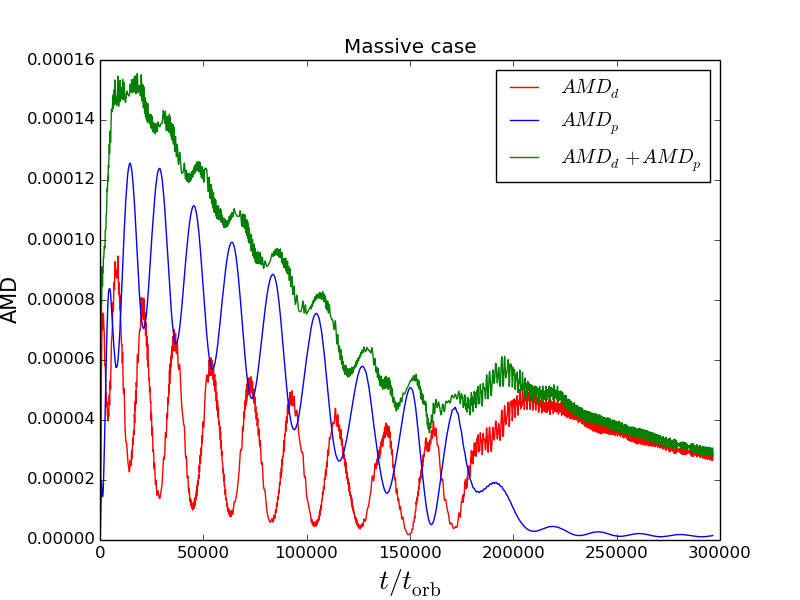}
\caption{AMD $A_{\rm p}$, $A_{\rm d}$, $A_{\rm tot}$ as a function of time for light (left panel) and massive (right panel) case. }
\label{fig:AMD}
\end{center}
\end{figure*}

Another interesting quantity that is useful for the interpretation of the results presented here is the AMD (equations (\ref{AMDeq}) and (\ref{AMDeqP}), Fig. \ref{fig:AMD}).
The evolution of the AMD reflects the evolution of the eccentricity since $A \propto e^2$ in the limit of low eccentricity (Eq.
(\ref{AMDappr})). The total AMD $A_{\rm tot}=A_{\rm p}+A_{\rm d}$ in the light case starts growing when the system evolves to the aligned configuration, while in the massive case it decreases up to the end of the simulation. 
However the most interesting feature of the oscillations observed in the eccentricity is that their amplitude is AMD conserving (see Fig. \ref{fig:AMD}): the overall amount of AMD changes both in the planet and in the disc, but the amplitude of the oscillations is such that the $A_{\rm tot}$ is conserved during one oscillation. The implication of this is that the periodic oscillations only {\it exchange} angular momentum between the planet and the disc: for a given amount of angular momentum exchanged the amplitude of the fluctuations is fixed by the orbital properties of the planet and the disc. 

As we will see, most of the features that we observed in these simulations (rapid initial exponential growth of eccentricity, long-term periodic oscillations, rigid precession of the pericentre phase, linear growth or decrease of the eccentricity at late times) can be interpreted in terms of either a classical linear theory of a 2-planet system \citep{murray1999}, or its extension to planet-disc interactions \citep{teyssandier2016}. 

We devote section \ref{sec:interpretation} and \ref{sec:toymodelinterp} to this interpretation.

\section{Interpretation of the results}\label{sec:interpretation}

Before trying to interpret the complex evolution revealed by the hydrodynamical simulations,  it is instructive to build up a qualitative picture of the evolution of dynamically coupled eccentric planet/disc systems.

The structure of an eccentric planet-disc system can in principal be described as a superposition of rigidly precessing normal modes, each
of which is characterised by its disc eccentricity profile $e(R)$ (normalised to planet eccentricity), growth rate $\gamma$, precession rate $\omega$ and angular offset $\Delta\Phi$ between the line of apses of the disc and planet \citep{teyssandier2016}.

The rate of precession $\omega$ is set by many different contributions. Among them, purely secular gravitational planet-disc interaction is expected to cause the prograde precession of the pericentre phase of both planet and disc, while pressure effects in discs with standard pressure profiles ($dP/dR<0$, where $P$ is the pressure radial profile) are expected to cause the retrograde precession of the pericentre phase (it can be shown that the precession rate observed in Fig. \ref{fig:phase} are consistent with that predicted by \citealp{teyssandier2016}). 

Mode growth is generically driven by resonances. With standard disc parameters, eccentric Lindblad resonances allow the disc and planet eccentricity to grow, while eccentric corotation resonances cause the eccentricity to decrease \citep{goldreich1980,goldreich2003,ogilvie2003}. The growth or damping of the planet eccentricity thus depends on the balance between these two opposite effects. \citet{goldreich1980} showed that if the planet does not perturb significantly the surface density of the disc, the corotation torque slightly exceeds the Lindblad one, damping the planet eccentricity toward circular orbits. In contrast, if the planet carves a sufficiently deep cavity or gap around the planet (at least a factor $\sim 10^{-3}$, \citealp{duffell2015}), and if no material replenishes the corotation region, the corotation torque saturates \citep{goldreich2003,ogilvie2003} so that dominance of the Lindblad resonances causes the eccentricity to grow. 
In addition to the effect of eccentric co-rotation resonances, the disc viscosity is also expected to circularize the gas orbits in the disc. Putting all this together implies that the disc eccentricity evolution can be expected to depend on the planet-star mass ratio,  planet eccentricity, pressure and disc viscosity \citep{artymowicz1994,crida2006}.

Each of the aforementioned effects have been included in the derivation by \citet{teyssandier2016} of the eigenmodes of an eccentric fluid disc, both with and without the inclusion of a planet.
In this formalism, the real part of the eigen-values associated to each eigen-mode corresponds to the precession rate.

\subsection{Case of no mode damping or driving}

We first consider the case where mode damping and pumping can be neglected. This implies the absence of resonant interactions and viscous effects and therefore means that the gravitational influence of the planet on the disc is mediated by the secular interaction, i.e. the response of the disc to the zero frequency ($\Omega_{\rm m} = 0$) component of the Fourier decomposition of the acceleration induced by the  planet.

The secular interaction can be visualised as being the response of the disc to an elliptical ring of material representing the  time average of the planet's mass distribution around its orbit. If the system is in a single mode the amplitude of disc and planet eccentricity is constant in time and the entire system undergoes rigid precession at a constant rate.

If however the system exists in a superposition of modes, each with characteristic eccentricity profile and precession rate, the net eccentricity of both the planet and the disc undergoes cyclical variations, that correspond to the beats of the fundamental modes, depending on the instantaneous phase relationship of the various modes.
The varying eccentricity of both planet and disc result in an exchange of angular momentum between the two components. Since for Fourier mode with frequency $\Omega_{\rm m}$ the relationship between energy exchange and angular momentum exchange is given by $\Delta E = \Omega_{\rm m} \Delta L$, it follows that the secular interaction involves zero energy exchange between planet and disc (recalling that $\Omega_{\rm m}=0$ for secular interaction). It is therefore convenient to consider the interaction in terms of the angular momentum deficit (AMD) defined in Eq. (\ref{AMDeq}) and (\ref{AMDeqP}).

The differential precession of an ensemble of modes results in a variation of the AMD of disc and planet at constant energy. Total angular momentum conservation requires that the total AMD of the planet plus disc is constant.

We have seen that the eccentricity and AMD variations of the simulated planet-disc system can indeed be described in terms
of such fluctuations on which slower long term trends in mode amplitude resulting from net pumping/damping are superposed. 

The fact that the oscillatory behaviour is close to being sinusoidal suggests  that the evolution can be understood in terms of the superposition of two dominant modes. We will find that we can gain significant qualitative insight into the behaviour of the system by considering the analogue problem of the secular interaction between two {\it point masses} for which (given the number of degrees of freedom in the system) there are just two modes (as known from textbook studies of celestial mechanics \citealp{murray1999}). We however emphasise that we do not necessarily expect the mode structure to be the same in the case of the fluid disc and will indeed find that - whereas the modes in the two-planet case both undergo prograde precession - the role of pressure within the disc can induce retrograde precession in one of the modes. Nevertheless, we will find that a heuristic understanding of the nature of the two modes in the point mass case will be extremely useful in guiding our interpretation of the simulations.

\subsection{Case of secular interaction between two point masses}
\label{sec:toymodel}
In this section we describe a toy model accounting only for secular contributions to the eccentricity equations. We aim to give a simplified description of the coupled evolution of planet and disc in order to interpret some features of the planet and disc eccentricity and pericentre phase evolution discussed in Sec. \ref{sec:results}.

\citet{teyssandier2017} predict a ``rigid'' evolution of the eigenmodes, leading us to expect that in general the eccentricity radial profile evolves rigidly as $e_{\rm d}(t,R)=e_{0}(R)h(t)$, where $h(t)$ is a generic function of the time only, and that the pericentre longitude does not depend on the radius $\Phi_{\rm d}(R,t)\equiv \Phi_{\rm d}(t)$. We thus expect that a simplified description of the evolution of the system can be obtained by replacing the disc with a virtual planet, adding some terms to account for the disc eccentricity pumping and damping effects. 
In fact this approach consists in modelling the planet-disc interaction as a planet-planet interaction where the outer planet is a virtual mass with disc averaged orbital characteristics: semi-major axis $a_{\rm d}$, longitude of pericentre $\Phi_{\rm d}$, mass $M_{\rm d}$ and eccentricity $e_{\rm d}$. We will use $a_{\rm p}$, $\Phi_{\rm p}$, $M_{\rm p}$ and $e_{\rm p}$ to refer to the actual planet instead. 

It is important to bear in mind that such a description will not be quantitatively correct for two main reasons: a) this approach intrinsically neglects pressure (which we have seen in the previous section to have a role in determining the precession rate) and b) the approximation of a disc of nested ellipses by an equivalent point mass particle forces us to reduce local quantities such as the density or the disc eccentricity to equivalent global quantities without a well defined prescription. Moreover such an approach does not of course include the additional effects of viscous damping and driving of eccentricity at resonances, which would need to be added {\it ad hoc}.

In the following equations we will use the following notation
\begin{equation}
E_j=|E_j|e^{i\Phi_j},\ j=\{{\rm p,d}\},
\end{equation}
where $|E_j|=e_j$ is the ``physical'' eccentricity and $\Phi_j$ is its pericentre phase, the subscripts $p$ and $d$ refer to the {\it planet} and {\it disc-``virtual'' planet}. This formalism allows us to write one single set of equations for both eccentricity and pericentre phase.

We follow the Hamiltonian approach given by \citet{zhang2013}, in which the gravitational potential produced by the two components (in our case the real planet and the disc virtual planet) of the system is expanded up to the second order in $e_{\rm p}$ and $e_{\rm d}$. The equations ruling the evolution of the complex eccentricities $E_{\rm p}$ and $E_{\rm d}$ have the form
\begin{equation}
 \left(\begin{array}{c}
 \dot E_{\rm p}\\
 \dot E_{\rm d}
 \end{array}\right)=\bm{\mathsf{M}}
   \cdot \left(\begin{array}{c}
   E_{\rm p}\\
   E_{\rm d}
   \end{array}\right),\label{ecceq}
\end{equation}
where the notation $\dot E_i$ indicates the time derivative, while the complex matrix $\bm{\mathsf{M}}$ reads
\begin{equation}
\bm{\mathsf{M}}=i\Omega_{\rm sec}
   \left(
   \begin{array}{cc}
      q & -q\beta \\[10pt]
    -\sqrt{\alpha}\beta &  \sqrt{\alpha}
   \end{array}
   \right),\label{Mim}
\end{equation}
where $\alpha=a_{\rm p}/a_{\rm d}$, $q=M_{\rm d}/M_{\rm p}$, $\beta=b^{(2)}_{3/2}(\alpha)/b^{(1)}_{3/2}(\alpha)$, where $b^{(n)}_{3/2}(\alpha)$ is the $n$-th Laplace coefficient 
\begin{equation}
b_{3/2}^{(n)}(\alpha)= \frac{1}{\pi}\int^{2\pi}_0\frac{\cos(n\theta)}{(1-2\alpha+\alpha^2)^{3/2}}d\theta.
\end{equation}
For $\alpha\ll 1$, $\beta\approx 5\alpha/4$ since $b^{(1)}_{3/2}(\alpha)\approx 3\alpha$ and $b^{(2)}_{3/2}(\alpha)\approx 15\alpha^2/4$ \citep{murray1999}.
The matrix $\bm{\mathsf{M}}$ in Eq. (\ref{Mim}) is purely imaginary and accounts for the secular, non-dissipative planet-disc interaction. $\Omega_{\rm sec}$ is a real scaling parameter for the matrix and has the dimension of a frequency:
\begin{equation}
\Omega_{\rm sec}=\frac{1}{4}\Omega_{\rm p}\frac{M_{\rm p}}{M_\star}\alpha^2b^{(1)}_{3/2}(\alpha).\label{omegasec}
\end{equation}

The solutions to equation (\ref{ecceq}) is
\begin{equation}
   \left(\begin{array}{c}
   E_{\rm p}(t)\\
   E_{\rm d}(t)
   \end{array}\right)=C_1\left(\begin{array}{c}
 \eta_{\rm s}\\
 1 
\end{array}\right)e^{i\omega_{\rm s}t}+C_2\left(\begin{array}{c}
 \eta_{\rm f}\\
 1 
\end{array}
\right) e^{i\omega_{\rm f}t},\label{generalsolution1}
\end{equation}
where $C_1$ and $C_2$ are constants that depend on the initial conditions, $\omega_{\rm s,f}$ and $(\eta_{\rm s,f},1)$ are the complex eigen-values and complex eigen-vectors of $\bm{\mathsf{M}}$, respectively.


\begin{figure}
\begin{center}
\includegraphics[width=0.49\textwidth]{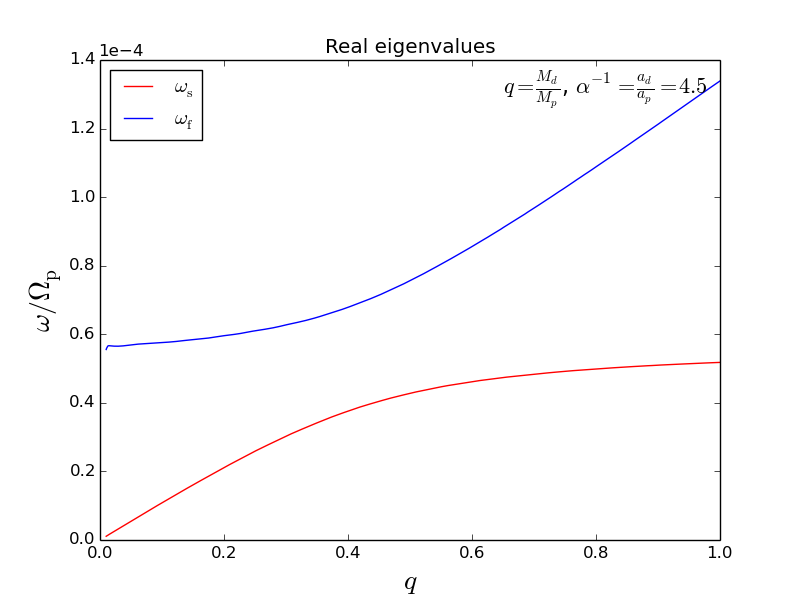}
\caption{Eigenvalues as a function of $q$ for fixed $\alpha^{-1}=4.5$. The blue curve represents $\omega_{\rm f}$ while the red one $\omega_{\rm s}$. The eigenvalues are expressed in units of the planet orbital frequency $\Omega_{\rm p}$, and represent the precession rate of the pericentre phase. }
\label{fig:eigv}
\end{center}
\end{figure}

\begin{figure*}
\begin{center}
\includegraphics[width=0.49\textwidth]{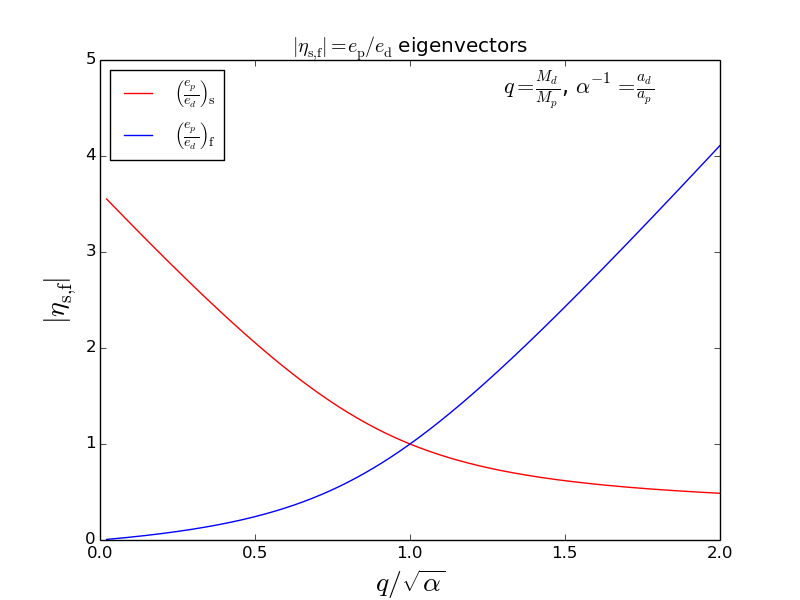}
\includegraphics[width=0.49\textwidth]{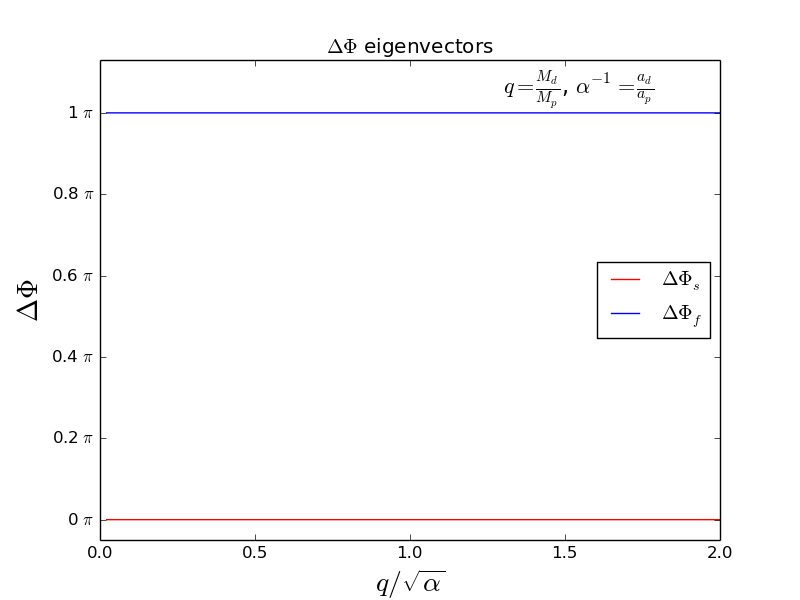}
\caption{$|\eta_{\rm s,f}|=e_{\rm p}/e_{\rm d}$ (left panel) and $\Delta\phi=\arg(\eta_{\rm f,s})$ (right panel) of eigenvectors as a function of $q/\sqrt{ \alpha}$. The blue curve refers to the fast mode the red curve refers to the slow one. It should be remembered that $e^{i\pi}=-1$. $\eta_{\rm s,f}$ represent the ratio between the planet and the disc eccentricity if just one mode is present. While $\Delta \Phi$ represents the pericentre phase difference between the planet and the disc if just one mode is present.}
\label{fig:eigvec}
\end{center}
\end{figure*}

With reference to Eq. (\ref{generalsolution1}), the eigen-values $\omega_{\rm s,f}$ of the matrix $\bm{\mathsf{M}}$ in Eq. \ref{Mim} are
\begin{equation}
\omega_{\rm s,f}= \frac{1}{2}\Omega_{\rm sec} \left(q+\sqrt\alpha\right)\mp \frac{1}{2}\Omega_{\rm sec}\sqrt{ \left(\sqrt\alpha-q\right){}^2+4 q\sqrt{\alpha}\beta^2}\label{varpi}
\end{equation}
The eigen-frequency $\omega$ in units of $\Omega_{\rm p}$, plotted for varying $q$ and fixed $\alpha$ in Fig. \ref{fig:eigv}, gives us information on the precession rate of the pericentre phase. From Eq. (\ref{varpi}) we can clearly see that $\omega_{\rm s}<\omega_{\rm f}$; for this reason we will refer to the $s$ mode as the ``slow mode'' and to the $f$ mode as the ``fast mode''. It is notable that $\Omega_{\rm sec}$ sets the timescale of the oscillations, it is independent of $q$ and scales as $\alpha^3$ for $\alpha << 1$ (equation (\ref{omegasec})). The individual precession frequencies and hence the beat frequency $\Delta \omega = \omega_{\rm f} - \omega_{\rm s}$ do however depend on $q$; in particular, the dependence on $q$ mostly affects the beat frequency $\Delta\omega=\omega_{\rm f}-\omega_{\rm s}$, which has a minimum when $q=\sqrt{\alpha}$ (see Fig. \ref{fig:eigv}). 

The components of the eigen-vectors of the matrix $\bm {\mathsf{M}}$ instead are
\begin{equation}
\eta_{\rm s,f}=    \frac{1}{2\sqrt{\alpha}\beta}( \sqrt{\alpha}-q)\pm \frac{1}{2\sqrt{\alpha}\beta}\sqrt{ \left(\sqrt{\alpha}-q\right){}^2+4 q\sqrt{\alpha}\beta^2}.\label{reigv}
\end{equation}
According to Eq. (\ref{generalsolution1}), the ratio of planet to disc eccentricity when only one of the modes is present is given by $|\eta|=e_{\rm p}/e_{\rm d}$.

It should be noticed that $\eta_{\rm f}<0$ for any parameter choice, while in contrast $\eta_s>0$.
This implies that the fast mode involves misalignment by $\pi$ between the pericentre phases of the planet and virtual planet (i.e. disc)
while the two orbits are aligned in the case of the slow mode. This is illustrated in the right hand panel of Fig. \ref{fig:eigvec}. It should be also noticed that for $q/\sqrt \alpha<1$ one has $|\eta_{\rm s}|>1$ and $|\eta_{\rm f}|<1$ (left panel of Fig. \ref{fig:eigvec}), while for $q/\sqrt \alpha>1$ one gets $|\eta_{\rm s}|<1$ and $|\eta_{\rm f}|>1$. The condition $q = \sqrt{\alpha}$ marks the condition that the two components have equal angular momentum if on a circular orbit\footnote{It should be noticed that the ratio $q/\sqrt{\alpha}$ is equivalent to the ratio $J_{\rm d,circ}/J_{\rm p,circ}=(M_{\rm d}\sqrt{GM_\ast a_{\rm d}})/(M_{\rm p}\sqrt{GM_\ast a_{\rm p}})$, and provides thus a measure of the relative contribution of the disc and the planet to the total amount of angular momentum of the system.\label{qsqrta}}.

The above inequalities imply that the component (i.e. planet or virtual planet) with the higher circular angular momentum will have the greater amplitude when the system is entirely in the slow mode ($C_2=0$), while the component with lower circular angular momentum will have greater amplitude when the system is entirely in the fast mode ($C_1=0$).
In the limit that the circular angular momenta of the two components is very different (i.e.: $q\rightarrow 0$), the fast mode becomes overwhelmingly dominated by the component with the smaller angular momentum and in this limit can be envisaged as the precession of a test particle in the combined potential of the central object and the potential generated by a circular ring of material at the location of the other `planet'. In this limit the slow mode is non precessing and has finite eccentricity contributions in both components. As $q$ tends to $\sqrt{\alpha}$ the modes become increasingly entwined in the sense that both modes contain comparable contributions in both components. We will discuss in the following sections how the evolution of the two simulations described here can be understood in terms of the different ratios of $q/\sqrt{\alpha}$ in the two cases, and thus$^{\ref{qsqrta}}$ of $J_{\rm d,circ}/J_{\rm p,circ}$.

Given the large number of simulations with fixed binaries in the literature \citep{dangelo2006,muller2013,duffell2015,thun2017}, we believe it is very interesting to notice that the case $q=0$ is the reference case for those simulations of a circumbinary disc surrounding a binary system (of any mass ratio) with fixed orbital parameters. A discussion of the instructive case $q=0$ can be found in Appendix \ref{caseq0}.

\subsection{Phenomenological implementation of pumping and damping terms}

The mere gravitational interaction we discussed in the previous section does not provide any mode evolution since it describes stationary modes. Nevertheless, we clearly observe instead in our simulations the growth and the decrease of the eccentricity at different stages of the simulations. This clearly implies that some modelling of this behaviour needs to be included in our simplified description.

\citet{zhang2013} treated the damped three body problem in order to show that the damping of the eccentricity of a hot-Jupiter operated by the tidal effects of the central-star can be slowed down if a second planet orbiting on an outer orbit is present. To do so, they added some real terms in the matrix $\bm{\mathsf{M}}$ in Eq. (\ref{ecceq}). The introduction of these terms introduces a complex component of the eigen-values which is responsible for the exponential damping (or pumping depending on the sign) of the mode. 

In our simulations, the resonances are initially very strong due to the presence of a large amount of material in the cavity region during the initial phases. For this reason the planet eccentricity in the massive case and the disc one in both cases grows very fast for $t\lesssim 10^4$ orbits following an exponential trend.  However, after this initial transient, the gas depletion in the cavity region leads to the saturation of the pumping mechanism, which is also associated with the attainment of a maximum value of the disc eccentricity. At later times the pumping/damping mechanisms are such to cause a linear increase/decrease of the eccentricity with time, in contrast with the exponential trend predicted by \citet{zhang2013} modelling to include resonant and viscous effects.

As described at the beginning of Sec. \ref{sec:interpretation}, the physical scaling of pumping and damping needs to account for viscous effect and the contribution of each single resonance.
The work by \citet{teyssandier2016} includes a detailed formulation to deal with pumping and damping effects. The viscous effects depend on the disc eccentricity radial gradient and on the viscosity prescription adopted; it provides damping of eccentricity for standard disc parameters. For the resonant interaction,  each individual resonance depends differently on several factors including: the disc density profile, the planet eccentricity and pressure effects, which vary significantly throughout the simulation.

In the light of these considerations, we conclude that our ability to model the eccentricity pumping/damping in the framework of the toy model is very limited. The main difficulty resides in identifying the dependence of the pumping/damping on global planet and disc properties.
We decide for this reason to include these effects in our model by prescribing a linear time evolution of the $C_1$ and $C_2$ parameters in Eq. (\ref{generalsolution1}), which becomes
\begin{equation}
   \left(\begin{array}{c}
   E_{\rm p}(t)\\
   E_{\rm d}(t)
   \end{array}\right)=C_1(t)\left(\begin{array}{c}
 \eta_{\rm s}\\
 1 
\end{array}\right)e^{i\omega_{\rm s}t}+C_2(t)\left(\begin{array}{c}
 \eta_{\rm f}\\
 1 
\end{array}
\right) e^{i\omega_{\rm f}t}.\label{generalsolution}
\end{equation}
We prescribe the time dependence of $C_1(t)$ and $C_2(t)$ to be in the form
\begin{align}
C_1(t)&=\max(C^0_1+\gamma_{\rm s}t,0.01)\label{pdump1}\\
C_2(t)&=\max(C^0_2+\gamma_{\rm f}t,0.01)\label{pdump2}
\end{align}
where $\gamma_{\rm s}>0$ and $\gamma_{\rm f}<0$ are pumping and damping rates with the dimension of frequencies. We keep a minimum value of $C_{1,2}=0.01$ to prevent $C_{1,2}$ from becoming negative, since in both cases oscillations are present up to the end of the simulations (indicating that both the modes maintain an amplitude $\neq 0$). The sign of $\gamma_{\rm s,f}$ is chosen on the basis of what we observe in our simulations, i.e. a transition from the fast to the slow mode. 

Solutions like eq. (\ref{generalsolution}) imply that the eccentricity $|E_{\rm p}|=e_{\rm p}$ and $|E_{\rm d}|=e_{\rm d}$ of planet and disc are
\begin{align}
|E_{\rm p}|&=\sqrt{C_1^2(t)\eta_{\rm s}^2+C_2^2(t)\eta_{\rm f}^2+2C_1(t)C_2(t)\eta_{\rm s}\eta_{\rm f}\cos(\Delta\omega t)},\label{eccp}\\
|E_{\rm d}|&=\sqrt{C_1^2(t)+C_2^2(t)+2C_1(t)C_2(t)\cos(\Delta\omega t)},\label{eccd}
\end{align}
where $\Delta\omega=\omega_{\rm f}-\omega_{\rm s}$. It becomes clear that the simulataneous presence of two eigen-modes produces in the eccentricity some typical oscillations with a periodicity equal to the beat frequency of the two precession rates $\Delta\omega$. 
 
The pericentre phase evolution of the planet $\Phi_{\rm p}(t)$ and of the disc $\Phi_{\rm d}(t)$ is given by
\begin{align}
\Phi_{\rm p}  &={\rm mod}\bigg\{\frac{\omega_{\rm s}}{2}t+\frac{\omega_{\rm f}}{2}t+\nonumber\\
        &\quad+\arg\left[(\eta_{\rm s}C_1(t)+\eta_{\rm f}C_2(t))\cos\left(\frac{\Delta\omega}{2}t\right)+\right.\label{phasep}\\
        &\qquad\quad\left.\left.i(\eta_{\rm f}C_2(t)-\eta_{\rm s}C_1(t))\sin\left(\frac{\Delta\omega}{2}t\right)\right],2\pi\right\},\nonumber\\
\Phi_{\rm d}  &={\rm mod}\bigg\{ \frac{\omega_{\rm s}}{2}t+\frac{\omega_{\rm f}}{2}t+\nonumber\\
&\quad+\arg\left[(C_1(t)+C_2(t))\cos\left(\frac{\Delta\omega}{2}t\right)+\right.\label{phased}\\
&\qquad\quad\left.\left.+i(C_2(t)-C_1(t))\sin\left(\frac{\Delta\omega}{2}t\right)\right],2\pi\right\}.\nonumber
\end{align}
However, more relevantly, for any complex number in the the form $E=\mathcal{A}e^{i\omega_{\rm s} t}+\mathcal{B}e^{i\omega_{\rm f} t}$, where $\mathcal A$ and $\mathcal B$ here represent the mode strength ($C_{1,2}(t)\eta_{\rm s,f}$ for the planet or $C_{1,2}$ for the disc), it can be shown that the phase $\arg(E)=\Phi$ can be approximated by
\begin{align}
\Phi &\approx\begin{dcases}
          \omega_{\rm s}t+\frac{\mathcal{B}}{\mathcal{A}}\sin(\Delta\omega t),& {\rm if}\, \mathcal{B}\ll \mathcal{A}\\
         \omega_{\rm f}t-\frac{\mathcal{A}}{\mathcal{B}}\sin(\Delta\omega t),& {\rm if}\, \mathcal{A}\ll \mathcal{B}
         \end{dcases},\label{phit}
\end{align}
implying that whether a component is predominantly precessing at the slow or fast rate (and what is the amplitude of superposed oscillations on this mean precession rate) is determined by the relative values of $\eta_{\rm s}C_1(t)$ and $\eta_{\rm f}C_2(t)$ for the planet solution, and by $C_1(t)$ and $C_2(t)$ for the disc one. The variation in time of these parameters implies that the system might experience a transition from the dominance of one mode to the other. This might occur at different times in the planet and in the disc depending on the absolute value of the eigen-vectors. It is important to stress that, for a given configuration, one component can be dominated by one mode while the other not. In fact, the ratio of the amplitude of the fast and the slow modes is given by
\begin{equation}
\mathcal{R}_{\rm p}=\frac{\eta_{\rm f}C_2}{\eta_{\rm s}C_1}
\end{equation}
and
\begin{equation}
\mathcal{R}_{\rm d}=\frac{C_2}{C_1}
\end{equation}
for the planet and the disc, respectively. Since $\eta_{\rm f}\neq\eta_{\rm s}$, the ratio of the two amplitudes can be different in the two components of the system. In particular, for small disc masses (small $q$), $\eta_{\rm f}\ll \eta_{\rm s}$ (see Fig. \ref{fig:eigvec}) and thus the planet can be in the slow mode while the disc resides in the fast mode. 

It is important at this stage to notice that the sign of the eigen-vectors sets the planet-disc configuration of the pericentre precession: in our formulation of the toy model we used eigen-vectors of the form $(\eta_{\rm s,f},1)$. For the slow mode, since $\eta_{\rm s}>0$, as we have seen in the previous section, both the disc and the planet component of the eigen-vector are positive; thus, with reference to Eq. (\ref{phit}), when both the planet and the disc satisfy $\mathcal B\ll \mathcal A$ they will precess at the slow precession rate with the disc and planet pericentres aligned. In contrast, for the fast mode, since $\eta_{\rm f}<0$, when both planet and disc complex eccentricities satisfy $\mathcal A\ll \mathcal B$ the planet and disc will precess at the fast rate with the pericentres anti-aligned. It follows that when the transition from the fast to the slow mode has been completed in both the planet and in the disc, we will observe also a transition from an anti-aligned to an aligned configuration of the planet-disc pericentre precession.

The initial values of $C_1^0$ and $C_2^0$ depend on the initial evolution of the system and, as previously said, cannot be predicted {\it a priori}. As mentioned in Sec. \ref{intro}, previous works found at short timescales a dependence of the growth rate and saturation value of the planet eccentricity on three main parameters: the planet mass, the disc mass and the initial value of the planet orbital eccentricity \citep{papaloizou2001,dangelo2006,dunhill2013,muller2013,duffell2015,thun2017}. Low mass planets ($M_{\rm p}\approx 1 \,M_{\rm J}$), with low levels of initial eccentricity and low disc masses have been observed to develop low eccentricities on the short timescales. In contrast, eccentric higher mass planets ($M_{\rm p}\approx 10 \,M_{\rm J}$), embedded in massive discs are more likely to show eccentricity growth during the initial phases of the simulation.

\section{Interpretation in the light of the toy model}\label{sec:toymodelinterp}

From Eq. (\ref{eccp}) and (\ref{eccd}) the relative amplitude of the oscillations between the planet and the disc is set by $\eta_{\rm s}\eta_{\rm f}=-q/\sqrt{\alpha}$. Since $\eta_{\rm s}\eta_{\rm f}<0$, the oscillations in the eccentricity between the planet and the disc are anti-phased, as can be noticed in Fig. \ref{fig:ecc}. This enables the conservation of the total AMD across the time of one oscillation as would be expected in pure planet-planet interaction in celestial mechanics (the non-conservation of the AMD on longer timescales is due to the pumping/damping effects). 

Fig. \ref{fig:phase} shows clearly that at the beginning of both simulations the disc and planet pericentre phases are precessing in an anti-aligned configuration ($\Delta\Phi\approx 180^\circ$). In the light case (left panel of Fig. \ref{fig:phase}), after $\approx 4\times 10^4$ orbits the planet pericentre phase decouples from the disc one and starts precessing at a much slower rate. The same conclusion can be reached regarding the massive case, but the anti-aligned configuration lasts for much longer and is apparently broken only after $t\gtrsim 2\times 10^5$ orbits.

In the light case this behaviour can be easily interpreted as the coexistence of the two evolving eigen-modes with positive precession rates $\omega_{\rm s,f}$ predicted in Eq. (\ref{generalsolution}), assuming that the following relationships between $C_1,\,C_2,\,\eta_{\rm s},\,\eta_{\rm f}$ hold:
\begin{align}
C_1\eta_{\rm s}<C_2\eta_{\rm f},&\quad \; {\rm if}\; t\lesssim 4\times 10^4\, t_{\rm orb},\\
C_1\eta_{\rm s} >C_2\eta_{\rm f},&\quad \; {\rm if}\; t\gtrsim 4\times 10^4\, t_{\rm orb},\\
C_1<C_2 ,&\quad \; {\rm if}\; t\lesssim 2\times 10^5\,t_{\rm orb},\\
C_1>C_2 ,&\quad \; {\rm if}\; t\gtrsim 2\times 10^5\,t_{\rm orb}.
\end{align}
Note that in the light case $q\ll\sqrt{\alpha}$ and thus $\eta_{\rm f}\ll \eta_{\rm s}$.

In the massive case, the same interpretation can be given but with different times delimiting the different stages in the modes evolution
\begin{align}
C_1\eta_{\rm s}<C_2\eta_{\rm f},&\quad \; {\rm if}\; t\lesssim2\times 10^5\, t_{\rm orb},\\
C_1\eta_{\rm s} >C_2\eta_{\rm f},&\quad \; {\rm if}\; t\gtrsim2\times 10^5\, t_{\rm orb},\\
C_1 <C_2 ,&\quad \; {\rm if}\; t\lesssim 2\times 10^5\,t_{\rm orb},\\
C_1 >C_2 ,&\quad \; {\rm if}\; t\gtrsim 2\times 10^5\,t_{\rm orb}.
\end{align}
Note that here $q\gtrsim \sqrt{\alpha}$ and thus $\eta_{\rm f}\gtrsim\eta_{\rm s}$.

The time at which the transition between the fast and slow mode occurs depends both on the conditions after the initial transient and on the parameters $q$ and $\alpha$ involved in the model.

Although we do not know {\it a priori} what are the conditions in the disc at the end of the initial transient stage, we can use the picture
of the mode structure outlined in Sec. \ref{sec:toymodel} in order to understand how the evolution of the two simulations differs on account of different values of $q$ (effective disc to planet mass ratio) on long timescales. The smaller the mass of the disc, the more angular momentum is concentrated in the planet, which means that the slow mode has a larger contribution from the planet eccentricity. As the fast mode is damped, therefore, the planet can be expected to make the transition to being predominantly in the slow mode at an earlier evolutionary phase in the light case than in the massive simulation. This is consistent with what we observe in our simulations:
in the light case the planet makes the transition to the slow precession mode at an earlier evolutionary stage (see Fig. \ref{fig:phase}).

\subsection{Evolution depending on the disc mass}
\begin{figure*}
\begin{center}
\includegraphics[width=0.49\textwidth]{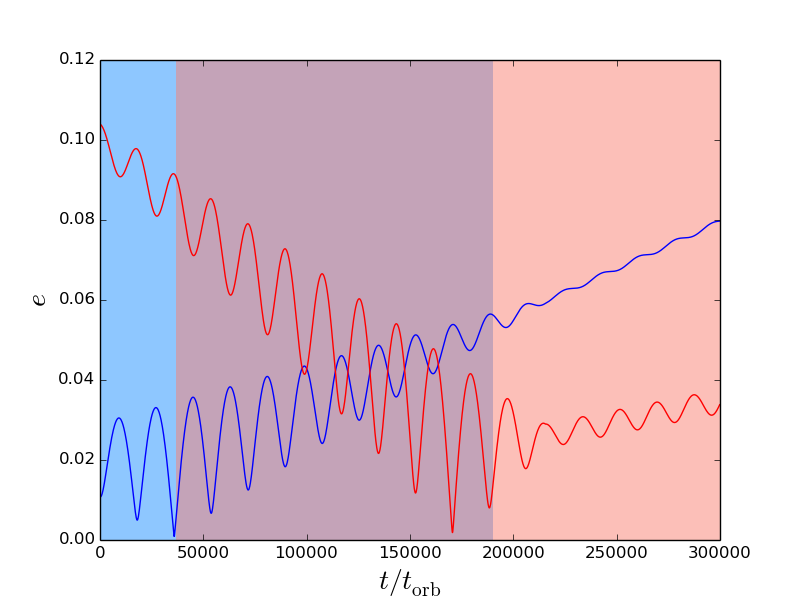}
\includegraphics[width=0.49\textwidth]{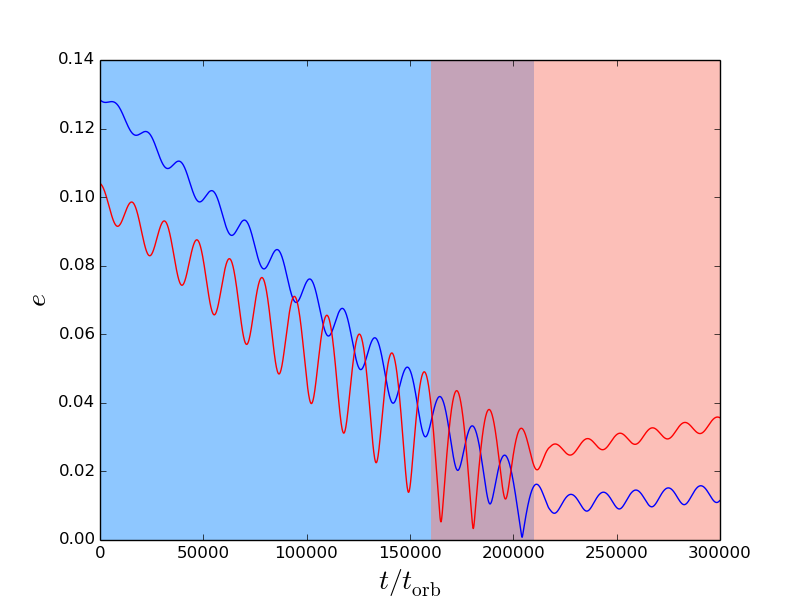}
\caption{Planet (blue curve) and disc (red curve) eccentricity using Eq. (\ref{eccp},\ref{eccd}). The left panel shows a reasonable choice of parameters for the light case, while the right panel a choice for the massive case. The summary of the parameters used can be found in Table \ref{tabpar}. 
The evolution of $C_2$ stops when $C_2=0.01$ is reached, as prescribed in Eq. (\ref{pdump1},\ref{pdump2}), in order to prevent $C_2$ from becoming negative.
Notice that the initial conditions $C_1^0$ and $C_2^0$ and pumping/damping coefficients $\gamma_{\rm s}$ and $\gamma_{\rm s}$ are the same in the two simulations, while $\omega_{\rm s}$ and $\omega_{\rm f}$ were chosen in order to reproduce the behaviour in the simulations.
The blue shaded area of the plots marks the time region where the fast $f$ mode is dominant in both the planet and the disc, causing the pericentres to precess at the fast rate in the anti-aligned configuration; the red shaded are marks the time region where the slow $s$ mode is dominant in both the planet and the disc, causing the pericentres to precess at the slow rate in the aligned configuration; the violet area marks the region where the slow mode is dominant in the planet but not in the disc, causing a decoupling of the precession rates.
These figures are not meant to reproduce precisely the eccentricity evolution in Fig. \ref{fig:ecc} but to show that a change in the values of the eigenvectors produced by a different $q$ (while keeping fixed all the other relevant parameters) can give rise to very different evolutionary path of the system. }
\label{fig:eccteor}
\end{center}
\end{figure*}

\begin{figure*}
\begin{center}
\includegraphics[width=0.49\textwidth]{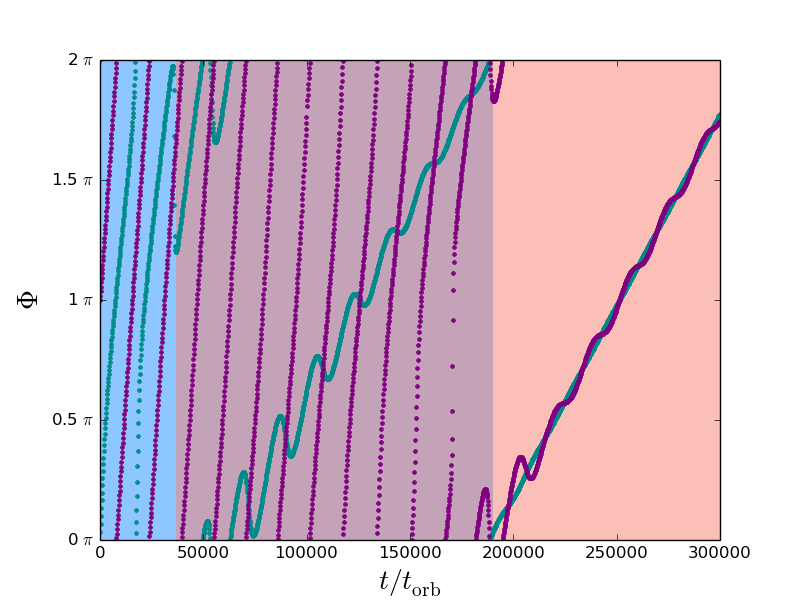}
\includegraphics[width=0.49\textwidth]{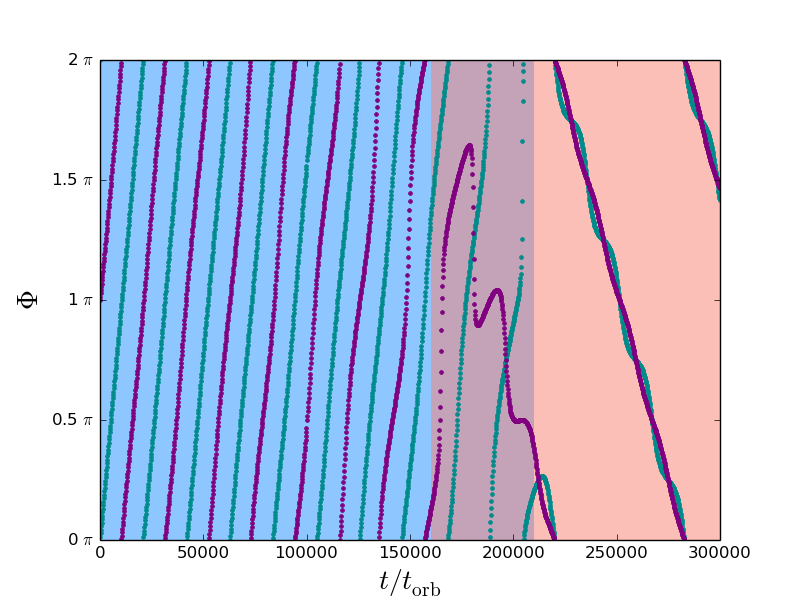}
\caption{Planet (blue curve) and disc (red curve) pericentre phase using Eq. (\ref{phasep},\ref{phased}). The left panel shows a reasonable choice of parameters for the light case, while the right panel a choice for the massive case. The summary of the parameters used can be found in Table \ref{tabpar}.
The evolution of $C_2$ stops when $C_2=0.01$ is reached, as prescribed in Eq. (\ref{pdump1},\ref{pdump2}), in order to prevent $C_2$ from becoming negative.
Notice that the initial conditions $C_1^0$ and $C_2^0$ and pumping/damping coefficients $\gamma_{\rm s}$ and $\gamma_{\rm s}$ are the same in the two simulations, while $\omega_{\rm s}$ and $\omega_{\rm f}$ were chosen in order to reproduce the behaviour in the simulations.
The blue shaded area of the plots marks the time region where the fast $f$ mode is dominant in both the planet and the disc, causing the pericentres to precess at the fast rate in the anti-aligned configuration; the red shaded are marks the time region where the slow $s$ mode is dominant in both the planet and the disc, causing the pericentres to precess at the slow rate in the aligned configuration; the violet area marks the region where the slow mode is dominant in the planet but not in the disc, causing a decoupling of the precession rates.
These figures are not meant to reproduce precisely the eccentricity evolution in Fig. \ref{fig:ecc} but to show that a change in the values of the eigenvectors produced by a different $q$ (while keeping fixed all the other relevant parameters) can give rise to very different evolutionary path of the system.} 
\label{fig:phaseteor}
\end{center}
\end{figure*}

\begin{table*}
\begin{center}
\begin{tabular}{c|cccccccccc}
\hline
 Simulation & $q$ & $\alpha$ & $C_1^0$ & $C_2^0$ & $\gamma_{\rm s}/ \Omega_{\rm p}$ & $\gamma_{\rm f}/ \Omega_{\rm p}$ & $\omega_{\rm s}/ \Omega_{\rm p}$ & $\omega_{\rm f}/ \Omega_{\rm p}$ & $\eta_{\rm s}$ & $\eta_{\rm f}$\\ \hline
light case & $ 0.2$ & $\approx 0.25$ & $0.004$ & $0.1$ & $1.60\times 10^{-8}$ & $-7.15\times 10^{-8}$ &  $7.95\times 10^{-6} $ & $6.35\times 10^{-5}$ & $2.3$ & $-0.2$ \\
massive case & $ 0.65$ & $\approx 0.25$ & $0.004$ & $0.1$ & $1.60\times 10^{-8}$ & $-7.15\times 10^{-8}$ &  $-1.60\times 10^{-5} $ & $6.35\times 10^{-5}$ & $0.4$ & $-1.3$ \\
\hline
\end{tabular}
\center
\caption{Summary of the parameters used to produce Fig. \ref{fig:eccteor} and \ref{fig:phaseteor}.}\label{tabpar}
\end{center}
\end{table*}

We now illustrate how the difference between the two simulations can be understood purely in terms of the dependence of the eigen-vectors on $q$. In Fig. \ref{fig:eccteor} and \ref{fig:phaseteor} we plot Eq. (\ref{eccp},\ref{eccd}) and Eq. (\ref{phasep},\ref{phased}) with a set of parameter choices for each simulation that qualitatively reproduce the main evolutionary features observed in our simulations (Fig. \ref{fig:ecc} and \ref{fig:phase}).

In both Fig. \ref{fig:eccteor} and \ref{fig:phaseteor} we prescribe the same pumping/damping prescription: a linearly decreasing fast mode ($\gamma_{\rm f}=-7.15\times 10^{-8}\,\Omega_{\rm p}$) and a growing slow mode ($\gamma_{\rm s}= 1.60\times 10^{-8}\,\Omega_{\rm p}$). We choose $\omega_{\rm s,f}$ to be consistent with those observed in the simulations. The summary of the parameters used in Fig. \ref{fig:ecc} and \ref{fig:phase} can be found in Table \ref{tabpar}.

This choice implies $\eta_{\rm s}/\eta_{\rm f}$ changes from being $> 1$ to $< 1$ between the simulations. We notice from Fig. \ref{fig:eigvec} that this implies that $q/\sqrt{\alpha}$ is respectively lower and greater than 1 (i.e. that the circular angular momentum is mainly in the planet in the light case and mainly in the disc in the massive case). We choose $C_1^0\ll C_2^0$ for both our simulations (see caption to Fig. \ref{fig:eccteor} and \ref{fig:phaseteor}). Noting that the ``density edge of the cavity'' is located at $R \approx 4.5$ in both simulations we adopt $\alpha \approx 0.25$ in both cases. The values of $q$ in the two simulations are $0.2$ and $0.65$ for the light and massive case, respectively.

Obviously our simulations are not supposed to share the initial conditions and mode pumping/damping rates, in contrast with what we prescribed.
However, we remark that Fig. \ref{fig:eccteor} and \ref{fig:phaseteor} are not meant to reproduce precisely the eccentricity evolution in Fig. \ref{fig:ecc}. We find it more instructive to show a comparison between the two regimes while keeping fixed all the other relevant parameters in order to highlight the role of the mass ratio $q$ in determining the evolution of the system.

We believe that these images show clearly that a change in the values of the eigenvectors produced by a different $q$ (while keeping fixed all the other relevant parameters) can give rise to very different evolutionary path of the system, despite our inability to model properly the evolution during the initial phases of the simulation. Furthermore, we see from Fig. \ref{fig:eccteor} and \ref{fig:phaseteor} that this simple parametrisation does an extraordinarily good job of reproducing the main features of the planet and disc eccentricity evolution on long timescales (compare with Fig. \ref{fig:ecc} and \ref{fig:phase}). 

In the light case the planet makes the transition to the slow mode significantly before the disc because the simulation is in the regime $q \ll \sqrt{\alpha}$ where the two eigenvectors are very different. In the massive case the value of $q/\sqrt{\alpha}$ is closer to $1$ and so the eigenvectors are more similar to each other. Consequently the disc and the planet follow more similar evolution of the eccentricity and indeed they make the transition to being predominantly in the slow mode at nearly the same time. By the end of the simulation both simulations are mainly in the slow mode and hence the ratio of eccentricities is given simply by the eigenvector of the slow mode.

\begin{figure}
\includegraphics[width=0.49\textwidth]{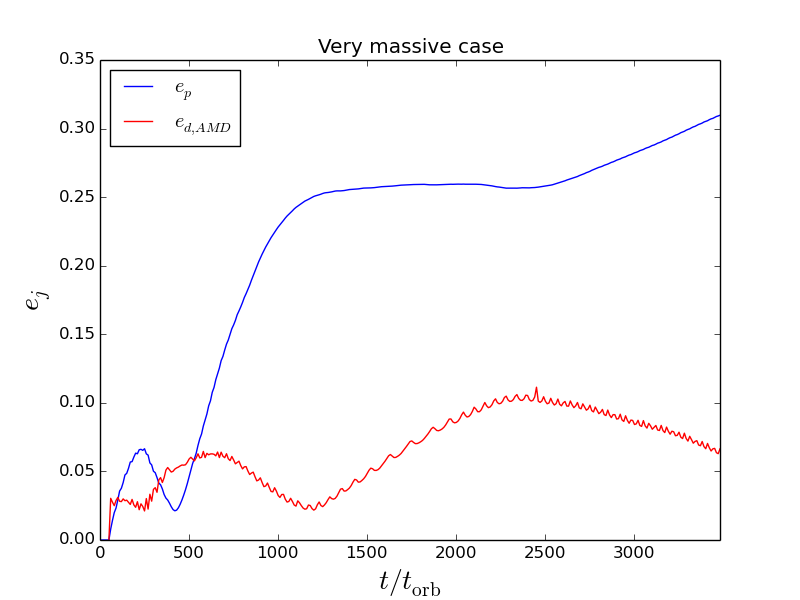}
\caption{Eccentricity evolution of planet (blue curve) and disc (red curve, AMD based) during the early stages of the simulation for a $q=2$ planet-disc system. This behaviour is consistent with what was observed in the previous work by \citet{papaloizou2001}. However our analysis suggests that this is a transient growth phase, and we expect the eccentricity to decay on a longer timescale}\label{eccVM}
\end{figure}

The main conclusion that can be drawn from these considerations, is that, {\it assuming} a damping of the fast mode, the system will end up in a configuration where $e_{\rm p}>e_{\rm d}$ for all those cases in which $q/\sqrt{\alpha}=J_{\rm d}/J_{\rm p}<1$ (light discs) and viceversa $e_{\rm p}<e_{\rm d}$ for all those cases in which $q/\sqrt{\alpha}>1$ (massive discs). As a consequence, under this assumption one should expect that low mass discs favour the growth of the eccentricity at long timescales.
We caution however that the assumption about the fast mode damping is tentative since it is based only on the two simulations we performed. A larger number of simulations is required to address the reliability of this assumption and the direction that further investigations should take.

It should be noted that our simulations show that higher disc masses can pump higher levels of planetary eccentricity at short timescales (as previously pointed out by \citealp{dunhill2013}). In addition, we have carried a third simulation with a much more massive disc ($q=2$), in order to compare with the previous results obtained by \citet{papaloizou2001}, finding that such a high disc mass allows the planet to reach eccentricities as high as $e=0.3$ during the first $3\times 10^3$ orbits (see Fig. \ref{eccVM}), consistently with what they previously found. This shows that the eccentricity of the planet acquired over hundreds of orbits does indeed increase with disc mass as one would naively expect but, if our analysis holds for such massive discs, we expect the eccentricity to decay on a longer timescale ($~10^5$ orbits).

We do not attempt to run this simulation for as many orbits as the two cases we presented in this paper. Due to the required computational resources, it is beyond the scope of this paper to further verify whether massive discs lead to more damping of the planet on long timescales, but future work should address this statement.

\subsection{Effects of disc viscosity and thickness}\label{discthickvisc}

Disc viscosity ($\nu$) and thickness ($H/R$) are expected to play a role in the eccentricity evolution of the planet-disc system. 

The effects of viscosity might act to either increase or damp the eccentricity. On the one hand, an increase in the disc viscosity implies stronger damping effects \citep{teyssandier2016}, providing thus an overall faster decrease of the system eccentricity. On the other hand, more viscous discs have smaller cavities, modifying the surface density at resonant locations and thus produce a stronger resonant interaction and faster evolution of the growing/decreasing trend. The pumping or damping nature of this latter effect depends on the type of resonances (Lindblad or co-rotation) that are strengthened. 

Regarding the disc thickness, a higher $H/R$ also provides a narrower gap, hence stronger resonant interaction again. Furthermore, the resonance width is broadened by pressure effects, and scales as $(H/R)^{2/3}$, so higher $H/R$ implies that resonances operate on a broader disc region, which also increases the growth/decrease rate. Higher $H/R$ implies a faster propagation speed of the mode, making the growth of a trapped slow mode more difficult in the inner regions \citep{teyssandier2016}. Finally, larger $H/R$ increases the effect of pressure, which drives retrograde precession. This opposes the gravitational secular interaction, which drives prograde precession. 

Finally, large eccentricity gradients imply large fluid relative velocities, that approach the sound speed when $Rde/dR \sim H/R$ and produce the crossing of fluid trajectories when $Rde/dR \sim 1$ \citep{ogilvie2001}, possibly inducing shocks and large pressure gradients that might limit the further growth of the disc eccentricity. Whether this implies that the planet eccentricity scales with $H/R$ \citep{duffell2015} still needs to be addressed with further work, although note that in our simulations we reach planet eccentricities $\approx 0.1$, well in excess of the value of $H/R\approx 0.036$.

A further exploration of the parameter space is required to address the dependence of the evolution on these parameters.

\subsection{Effects of the disc evolution}\label{sec:effectsdiscevo}

While our toy model reproduces the qualitative features we have highlighted so far, as already noticed in Sec. \ref{sec:results} a careful inspection of Fig. \ref{fig:ecc} reveals that the oscillation frequency of the eccentricity is not fixed in time. This effect cannot be captured within our modelling but it indicates that the viscous evolution of the disc has a role in determining the precession rate (eigen-frequencies) and modes relative strength (eigen-vectors): as the disc viscously spreads the effective $q$ and $\alpha$ change, causing an evolution of the eigenfrequencies. The accretion of material at the inner edge causes a further decrease of the effective $q$. In addition, it is very important to remember that the eccentricity values of the planet have an effect in determining the size and the density profile of the cavity edge \citep{artymowicz1994,thun2017} causing a variation of the effective $\alpha$.

As a consequence, it should be noticed that any system naturally evolves toward a situation where $q/\sqrt{\alpha}<1$ due to the progressive disc dispersal. This implies the existence of a period in which the planet eccentricity grows above that of the disc. However the growth of the eccentricity occurs on a very long time-scale, thus if the disc disperses too rapidly this final growth might not occur at all.

In the massive case the disc switches to the slow mode after $2\times 10^5$ orbits. The dominance of this single mode produces an abrupt change in the disc eccentricity and density profile. In particular, in this mode the resonant region is very depleted.
This not only explains the stalling of migration but also the fact that the eccentricity stops growing and is then subject just to very slow damping for the last $\sim 10^5$ orbits (likely due to viscous effects in the disc).

\section{Conclusions}\label{sec:conclusions}

We performed two long time scale 2D hydrodynamical simulations of a planet embedded in a gaseous disc using two different disc masses (light case and massive case), in order to study the long term evolution of both the planet and disc eccentricities.

The planet-disc interaction induces an eccentricity exchange between the planet and the disc in the form of periodic oscillations of both planet and disc eccentricity superimposed on a growing or decreasing trend depending on the disc mass. 

In the light disc case the planet eccentricity, after an apparent stalling of its evolution, grows linearly with time up to $e_{\rm p}\approx 0.12$ (reached after $3\times 10^5$ orbits). After $2\times 10^5$ orbits the growth appears to slow down, probably because of some saturation effects. The disc eccentricity rapidly reaches $e_{\rm d}\approx 0.1$ at the beginning of the simulation and then decreases linearly. At the end of the simulation the planet-to-disc eccentricity ratio is $(e_{\rm p}/e_{\rm d})_{\rm light}\approx 3$.

In the massive case instead the planet eccentricity grows exponentially up to $e_{\rm p}\approx 0.14$ during the initial phases of the simulation but then linearly decreases as a function of time. As in the low mass case, in the massive case the disc eccentricity grows rapidly during the initial phases of the simulation up to $e_{\rm d}\approx 0.1$ and then decreases linearly as a function of time. In this case the planet eccentricity exceeds the disc one ($e_{\rm p}/e_{\rm d}>1$) up to $t\approx 2\times 10^5$ orbits, when a rapid transition to $e_{\rm p}/e_{\rm d}\approx 0.3$ occurs. 

In our simulations we find that the planet eccentricity can reach values $e_{\rm p}\gtrsim 0.1$, well in excess of the value of $H/R\approx 0.036$ suggested by \citet{duffell2015} as the maximum value of the eccentricity. Furthermore, \citet{goldreich2003} and \citet{duffell2015} have argued for the need of a non-zero initial eccentricity of the planet to ensure the saturation of the corotation torque, and subsequent growth of the eccentricity. In our simulations, both the planet and the disc are initially on circular orbits, suggesting that corotation resonances might not necessarily need to be saturated to cause eccentricity growth.

We interpret the coupled evolution of the planet and disc eccentricity in terms of a superposition of secular modes whose relative amplitudes are slowly modified by resonant pumping and viscous damping. These modes are generically a rapidly precessing mode with anti-alignment between disc and planet pericentres and a slowly precessing aligned mode. The ratio of disc to planet eccentricity in each of these modes is generically $>1$ and $<1$, depending on the ratio between the planet and disc angular momenta. The dominance of the disc eccentricity in the fast mode implies that viscous damping preferentially damps the fast mode. The system thus ends up being completely in the slow mode at the end of the simulation but following very different evolutionary paths.

At the end of the low mass simulation, the planet is describable as being primarily in the slow mode which is growing very slowly whereas the disc is executing high frequency low amplitude libration about this slow precession. The high mass simulation has instead evolved to a situation where the planet is on an almost circular orbit with stalled migration and the eccentric disc undergoes retrograde precession due to pressure effects. At the end of this simulation the region encompassing the Lindblad resonances has been cleared of material so failing re-supply of this region (by accretion from the outer disc), the orbital evolution of the planet is stalled in an almost circular orbit.  

We provide a simplified toy model in which we treat the disc as a second ``virtual'' planet undergoing the secular interaction with the real one. This model depends only on two variables $q=M_{\rm d}/M_{\rm p}$ and $\alpha=a_{\rm p}/a_{\rm d}$, predicts the presence of two eigen-modes with respective eigen-values (setting the precession rate) and eigen-vectors (setting the ratio $e_{\rm p}/e_{\rm d}$).

Under the same initial conditions and pumping/damping prescription, we are able to qualitatively reproduce with our toy model the two very different evolutions of eccentricity and pericentre phase in the light case and massive one (Fig. \ref{fig:eccteor} and \ref{fig:phaseteor}, to be compared with Fig. \ref{fig:ecc} and \ref{fig:phase}). The different behaviour of the two simulations can be understood in terms of different eigenvectors characterizing the fast and slow precession modes. Indeed in Eq. (\ref{reigv}) (see also Fig. \ref{fig:eigvec}) predicts that, for the same $\alpha$, low values of $q$ produce a slow mode with $e_{\rm p}>e_{\rm d}$, high values of $q$ predicts $e_{\rm p}<e_{\rm d}$. This is in perfect agreement with the outcome of our simulations.

Simply requiring a different mass ratio between the ``virtual'' planet and the real one the toy model is able to explain:
\begin{itemize} 
\item Oscillations in the eccentricity.
\item Transition from the fast to the slow mode.
\item Aligned and anti-aligned configurations of the pericentre precession.
\item Faster transition to the slow mode of the planet in the light case than in the massive one.
\item Final values of the eccentricity when the system is fully in the slow mode.
\end{itemize}

Our model is not able to capture the intrinsic nature of the pumping and damping mechanism and thus it cannot be used as predictive tool to determine under which conditions the slow mode grows or decreases the eccentricity. However, we expect the disc thickness and viscosity to have a role in determining the intensity of the pumping and damping mechanisms as pointed out in Sec. \ref{discthickvisc}. The absolute mass of the planet might be relevant as well affecting the strength of the resonant planet-disc interaction. In contrast it provides larger cavities, and some resonances might be saturated.

Nevertherless, the model predicts some useful relationships that apparently hold between the planet and disc eccentricity depending on the disc-to-planet mass ratio and disc cavity size.

If we assume that the damping of the fast mode on very long timescales is a general result, massive discs appear to disfavour values of planet eccentricity higher than light discs at late stages of their evolution. For relatively high $M_{\rm d}/M_{\rm p}$, and disc density comparable with that produced by a $13\,M_{\rm J}$ planet, the system ends up in a slow mode configuration characterized by $e_{\rm p}/e_{\rm d}<1$. In contrast, light discs are expected to produce a slow mode with $e_{\rm p}/e_{\rm d}>1$, in fact favouring higher values of $e_{\rm p}$.
This goes in the opposite direction of what is often found in the literature, where high planet eccentricities have been observed to develop in presence of high disc masses on short timescales (\citealp{papaloizou2001}): they found that a $30\,M_{\rm J}$ planet can reach $e_{\rm p}\approx 0.3$ in less than $10^3\, t_{\rm orb}$ with $M_{\rm d}/M_{\rm p}\approx 2$. To support the results by \citet{papaloizou2001}, we report that some preliminary simulations we performed reached eccentricity values as high as $e_{\rm p}\approx 0.3$ for $M_{\rm d}/M_{\rm p}\approx 2$ after $t\approx 10^3 t_{\rm orb}$ (see Fig. \ref{eccVM}). If the damping of the fast mode at long timescales is confirmed to be a general feature, this implies in fact a reversal of the dependence on the disc mass of the eccentricity evolution on short timescales.

We caution that our last claim is tentative and supported only by two numerical simulations. A larger number of simulations exploring a wider range of disc masses is required in order to properly investigate the issue. In general, a more complete understanding of the origin of the mechanisms for pumping or damping of the eigen-modes is obviously required in order to make quantitative predictions about the eccentricity evolution at very late times and constitutes a possible follow up of this work. 

It should be also considered that in principle any planetary system passes through a phase in which the ratio $M_{\rm d}/M_{\rm p}\ll 1$, due to the progressive disc dispersal and accretion of material on the planet, which increases its mass.
If we for now adopt the assumption that planets attain larger eccentricities in the case of {\it lower} mass discs, then it raises some interesting possibilities about planet eccentricity evolution during disc dispersal. Whereas rapid dispersal (i.e. on a timescale $\ll 10^5$ orbits) would simply freeze the planet's eccentricity at its previous value, slow dispersal could instead cause the planet eccentricity to rise in the last stages. However, further investigations are required in order to understand how the secular eccentricity evolution is affected by substantial changes in the disc parameters throughout its lifetime.

Finally, we believe that this work demonstrates the importance of carrying out long timescale simulations when studying the planet-disc interaction in protoplanetary discs: both our simulations indeed undergo a complete inversion of the evolutionary trend on long timescales with respect to those shown in the initial phases. This however does not occur before $5\times 10^4$ orbits, which is beyond the timescales explored in previous simulations.

\section*{Acknowledgments}

We thank Gordon Ogilvie for useful comments and fruitful discussions. We thank the referee, Gennaro D'Angelo, for his constructive comments and useful suggestions that improved the manuscript.
This work has been supported by the DISCSIM project, grant agreement 341137 funded by the European Research Council under ERC-2013-ADG. JT acknowledges support from STFC through grant ST/L000636/1. 
This work used the Wilkes GPU cluster at the University of Cambridge High Performance Computing Service (http://www.hpc.cam.ac.uk/), provided by Dell Inc., NVIDIA and Mellanox, and part funded by STFC with industrial sponsorship from Rolls Royce and Mitsubishi Heavy Industries. We also thank the MIAPP for hosting us for the ``Protoplanetary Disks and Planet Formation and Evolution'' topical workshop held in Munich during June 2017. All the figures were generated with the python-based package matplotlib \citet{hunter2007}.
\nocite{*}
\bibliography{biblio}

\appendix
\onecolumn

\section{The limiting case $\lowercase{q}=0$}\label{caseq0}
 
To understand the physical meaning of the two eigen-modes, it is instructive to consider the limit $q=0$. This case is particularly interesting from a physical point of view because is the case in which the second outer planet has negligible mass. Thus it constitutes the reference situation for all those simulations in which the binary is kept fixed on its initial orbit with constant orbital eccentricity and semi-major axis \citep{dangelo2006,muller2013,duffell2015,thun2017}. The problem becomes in fact the classical restricted three body problem in which a test particle orbits a binary object. 

To further simplify the equations we assume also $\alpha\ll 1$, so that $\beta\approx 5/4 \alpha$ \citep{murray1999}. In this limit the eigen-frequencies reads
\begin{align}
\omega_{s0} &=0,\\
\omega_{f0}&=\Omega_{\rm sec}\sqrt{\alpha},\label{fastf}
\end{align} 
while the non-unit component of the eigenvectors $\eta$ reads
\begin{align}
\eta_{s0} &=\frac{4}{5\alpha},\\
 \eta_{f0}&=0.
\end{align}

With reference to Eq. (\ref{generalsolution1}), these values for eigen-frequencies and eigen-vectors imply the following two limiting cases: first, the system is completely in the fast $f$ mode, the inner planet ($M_{\rm p}$) has $e_{\rm p}=0$ while the outer planet (for this case $M_{\rm d}=0$) orbits with arbitrary eccentricity with a pericentre precession frequency given by Eq. \ref{fastf}, no oscillations of the outer-planet eccentricity are observed. Second, the system is completely in the slow $s$ mode, the inner planet has an eccentricity $e_{\rm p}=C_1\eta_{s0}\neq 0$ while the outer one orbiting with an eccentricity $e_{\rm d}=e_{\rm p}/\eta_{\rm s0}=C_1$, its pericentre phase does not experience any form of precession, no oscillations of the outer-planet eccentricity are observed. In both cases, obviously the inner planet pericentre phase do not experience any precession since in this limit it does not feel the the presence of the outer planet at all. 

In a mixed situation (the system is both in the slow and in the fast mode, i.e. $e_{\rm p}\neq 0,\;e_{\rm d}\neq e_{\rm p}/\eta_{s0}$)  if $e_{\rm d}>e_{\rm p}/\eta_{\rm s}$ the outer planet will experience a complete precession (the pericentre phase of the outer planet will complete a revolution of $360^\circ$ around the central star) with a precession rate given by \ref{fastf}. In contrast if $e_{\rm d}<e_{\rm p}/\eta_{s0}$, the outer planet librates around the pericentre phase of the inner planet spanning a range of phases that becomes progressively smaller as $e_{\rm d}\rightarrow e_{\rm p}/\eta_{s0}$.

We can further expand $\omega_{f0}$ in Eq. \ref{fastf} expliciting $\Omega_{\rm sec}$
\begin{equation}
\omega_{f0}=\frac{3}{4}\alpha^{7/2} \Omega_{\rm p}\frac{M_{\rm p}}{M_\star},\label{omefqzero}
\end{equation}
associated to a precession period $t_{\rm prec}$ of the outer planet pericentre phase given by
\begin{equation}
t_ {\rm prec}=\frac{4}{3}\alpha^{-7/2}\left(\frac{M_{\rm p}}{M_\star}\right)^{-1}t_{\rm orb},
\end{equation}
which is perfectly consistent with the precession rate predicted by \citet{moriwaki2004} for the restricted three body problem apart from higher order corrections in $e$ in the expansion of the perturbing potential and with the interpretation of the precession frequency in the fixed planet simulations in \citet{thun2017} (equivalent to the $q=0$ case).
Furthermore, $e_{\rm p}/\eta_{s0}$ has the equivalent role of $e_{\rm forced}$ in \citet{moriwaki2004} and also in this case the two expressions are perfectly consistent apart from higher order corrections in $e$ in the expansion of the perturbing potential.

\section{Phase evolution}\label{phasevo}
In this section we will present some analytical approximations of Eq. (\ref{phasep}) and (\ref{phased}) through which it will be possible to obtain Eq. (\ref{phit}). In order to simplify the notation, we will refer to the modulus part of the eigen-modes as $\mathcal A$ and $\mathcal B$, implying that $\mathcal{A}=\eta_{\rm s}C_1(t)$ and $B=\eta_{\rm f}C_2(t)$ for the planet equations, and that $\mathcal A=C_1(t)$ and $\mathcal B=C_2(t)$ for the disc ones. With this simplifying substitution Eq. (\ref{generalsolution}) reads
\begin{equation}
|E|e^{i\Phi }=\mathcal Ae^{i\omega_{\rm s}t}+Be^{i\omega_{\rm f}t},
\end{equation}
which can be restated also as:
\begin{equation}
|E|e^{i\Phi }=e^{i\frac{\omega_{\rm s}+\omega_{\rm f}}{2}t}\underbrace{(\mathcal A+ \mathcal B)\left[\cos\left(\frac{\omega_{\rm f}-\omega_{\rm s}}{2}t\right)+i\frac{\mathcal B-\mathcal{A}}{\mathcal{A}+\mathcal B}\sin\left(\frac{\omega_{\rm f}-\omega_{\rm s}}{2}t\right)\right]}_{(\mathcal{A}+\mathcal B)\cos\left(\frac{\Delta\omega}{2}t\right)+i(\mathcal B-\mathcal{A})\sin\left(\frac{\Delta\omega}{2}t\right)\,=\,\mathcal{A}e^{-i\frac{\Delta\omega}{2}t}+\mathcal B e^{i\frac{\Delta\omega}{2}t}}.
\end{equation}
We can obtain the pericentre phase $\Phi $ applying the logarithm on both sides and taking only the imaginary part:
\begin{equation}
\Phi =\arg(|E|e^{i\Phi})=\Im\left[\log\left(Ae^{i\omega_{\rm s}t}+Be^{i\omega_{\rm f}t}\right)\right]
\end{equation}
which gives\footnote{Obtained exploiting $\log x=1/2\log x +1/2\log x$.}:
\begin{equation}
\Phi  = \frac{\omega_{\rm s}}{2}t+\frac{\omega_{\rm f}}{2}t+\Im\left\{\log\left[\cos\left(\frac{\omega_{\rm f}-\omega_{\rm s}}{2}t\right)+i\frac{\mathcal B-\mathcal A}{\mathcal A+\mathcal B}\sin\left(\frac{\omega_{\rm f}-\omega_{\rm s}}{2}t\right)\right]\right\}.\label{appendix:notappr}
\end{equation}
Expanding Eq. (\ref{appendix:notappr}) to the first order in $\mathcal A/\mathcal B\rightarrow \{0,1,\infty\}$ one then gets
\begin{align}
 \Phi  &\approx\begin{dcases}
          \omega_{\rm s}t+\overbrace{\Im\left[\log\left(1+\frac{\mathcal B}{\mathcal A}e^{i( \omega_{\rm f}- \omega_{\rm s})t}\right)\right]}^{\approx\frac{\mathcal B}{\mathcal A}\sin(\Delta\omega t)},& {\rm if}\, \mathcal B\ll \mathcal A\\
         \frac{\omega_{\rm s}}{2}t+\frac{\omega_{\rm f}}{2}t+  
         \frac{ \mathcal B^2-\mathcal A^2}{4 \mathcal A \mathcal B}\tan \left(\frac{\Delta\omega t}{2}\right),
         & {\rm if}\, \mathcal A\sim \mathcal B \\
         \omega_{\rm f}t+\underbrace{\Im\left[\log\left(1+\frac{\mathcal A}{\mathcal B}e^{-i( \omega_{\rm f}- \omega_{\rm s})t}\right)\right]}_{\approx-\frac{\mathcal A}{\mathcal B}\sin(\Delta\omega t)},& {\rm if}\, \mathcal A\ll \mathcal B
         \end{dcases},\label{appendix:phit}
\end{align}
where $\Delta\omega=\omega_{\rm f}-\omega_{\rm s}$.

\label{lastpage}

\end{document}